\documentclass[reprint,prb,aps,amsmath,amssymb,superscriptaddress,nofootinbib]{revtex4-2}

\usepackage{graphicx}
\usepackage{dcolumn}
\usepackage{bm}
\usepackage{xcolor}
\usepackage[english]{babel}
\usepackage{fontspec}
\usepackage[
colorlinks, 
citecolor=blue, 
urlcolor=blue, 
linkcolor=red,breaklinks]{hyperref}
\usepackage{orcidlink}
\renewcommand{\vec}{\boldsymbol}
\newcommand{\vk}{\vec{k}}

\newcommand{\vd}{\vec{\delta}}
\newcommand{\g}{\gamma}
\begin{document}

\preprint{APS/123-QED}

\title{Electronic structure and optical signatures of highly-doped graphene}

\author{Sa\'ul Antonio Herrera-Gonz\'alez
\orcidlink{0000-0001-7404-9630}
}
\email{sherreragonzalez@flatironinstitute.org}
\affiliation{Depto. de Sistemas Complejos, Instituto de F\'isica, UNAM, Ciudad Universitaria, 04510 Ciudad de M\'exico, Mexico.}
\affiliation{Center for Computational Quantum Physics, Flatiron Institute, 162 5th Avenue, New York, NY 10010, USA}
\author{Guillermo Parra-Mart\'inez
\orcidlink{0000-0002-1595-2042}
}
\affiliation{IMDEA Nanoscience, C/ Faraday 9, 28049 Madrid, Spain}
\author{Francisco Guinea
\orcidlink{0000-0001-5915-5427}
}
\affiliation{IMDEA Nanoscience, C/ Faraday 9, 28049 Madrid, Spain}
\affiliation{Donostia International Physics Center, Paseo Manuel de Lardiz\'{a}bal 4, 20018 San Sebastián, Spain}
\author{Jose Angel Silva-Guill\'en
\orcidlink{0000-0002-0483-5334}
}
\email{joseangel.silva@imdea.org}
\affiliation{IMDEA Nanoscience, C/ Faraday 9, 28049 Madrid, Spain}
\author{Pierre A. Pantale\'on
\orcidlink{0000-0003-1709-7868}
}
\email{pierre.pantaleon@imdea.org}
\affiliation{IMDEA Nanoscience, C/ Faraday 9, 28049 Madrid, Spain}
\author{Gerardo G. Naumis
\orcidlink{0000-0002-1338-1522}
}
\affiliation{Depto. de Sistemas Complejos, Instituto de F\'isica, UNAM, Ciudad Universitaria, 04510 Ciudad de M\'exico, Mexico.}

\date{\today}

\begin{abstract}
Heavily doping graphene by intercalation can raise its Fermi level near an extended van Hove singularity, potentially inducing correlated electronic phases. Intercalation also modifies the band structure: dopants may hybridize with carbon orbitals and order into $\sqrt{3}\times\sqrt{3}$ or $2\times2$ superstructures, introducing periodic potentials that fold the graphene $\pi$ bands. Angle-resolved photoemission spectroscopy further shows a pronounced flattening of the conduction band near the M points, producing higher-order van Hove singularities. These effects depend strongly on the dopant species and substrate, with implications for both many-body physics and transport. We construct effective tight-binding models that incorporate dopant ordering, carbon-dopant hybridization, and $\pi$-band renormalization. Model parameters are obtained from density functional theory and reproduce dispersions observed in photoemission experiments. Using these models, we compute the optical conductivity and identify characteristic signatures associated with dopant ordering and hybridization. Our results provide a framework to interpret experimental spectra and to probe the superlattice symmetry of highly doped monolayer graphene.
\end{abstract}

\maketitle

\section{Introduction}

The discovery of superconductivity and correlated insulating states in graphene stacks \cite{Cao2018Correlated,Cao2018Unconventional,Lu2019Superconductors} has led to an increased interest in the many-body physics of graphene and other two-dimensional (2D) materials. Because their properties are highly susceptible to proximity effects, there are multiple approaches for enhancing their electronic interactions. While stacking multilayers with a relative twist and applying electric fields have been successfully employed strategies for engineering flat electronic bands in these materials, other proposals include the use of strain \cite{Mao2020Evidence}, magnetic fields \cite{Zhou2021SuperRTG,Zhou2022Isospin}, and alignment with nearly-commensurate substrates \cite{Qiangsheng2022Dirac}. In particular, soon after the discovery of graphene, it was suggested that heavy electron doping could be a promising strategy to enhance its electronic interactions and realize many-body phases \cite{Black2007Resonating,Uchoa2007Superconducting,Gonzalez2008Kohn,Nandkishore2012Chiral}.

Interaction effects are generally expected to get amplified in a system with a large electronic density of states (DOS) near the Fermi level $E_F$. While close to charge neutrality the DOS of graphene-like honeycomb lattices is vanishingly small, at higher energies they host a saddle van Hove (VH) point at each M point in the Brillouin zone (BZ), leading to a diverging DOS. As a consequence, it was predicted that many-body phases could be induced in graphene by shifting the Fermi level close to the VH singularity via electron doping. The predicted phase diagram in this high-doping regime includes different types of superconductivity, charge density waves and exotic magnetic states, among other phases \cite{Nandkishore2012Chiral,Kiesel2012Competing,Classen2020Competing}.

Experimentally, it has been reported that doping graphene to such levels can substantially modify its electronic structure. 
The electron density required for $E_F$ to approach the VH point is in the range of $n \sim 10^{14}$ cm$^{-2}$ \cite{McChesney2010Extended,Efetov2014Towards,Rosenzweig2020Overdoping}. This density is not achievable via the usual electric field effect due to  the dielectric breakdown of typical gates at densities above $n\sim 10^{12}$ cm$^{-2}$ \cite{Efetov2014Towards,Efetov2010Controlling}. Instead, achieving the required electron densities requires chemical doping of graphene, most commonly done by intercalation and adsorption of dopant atoms \cite{McChesney2010Extended,Rosenzweig2019Tuning,Rosenzweig2020Overdoping}. The resulting high charge density and the interactions with dopants can significantly renormalize the electronic structure of graphene.

\begin{figure*}
\includegraphics[width=1\textwidth]{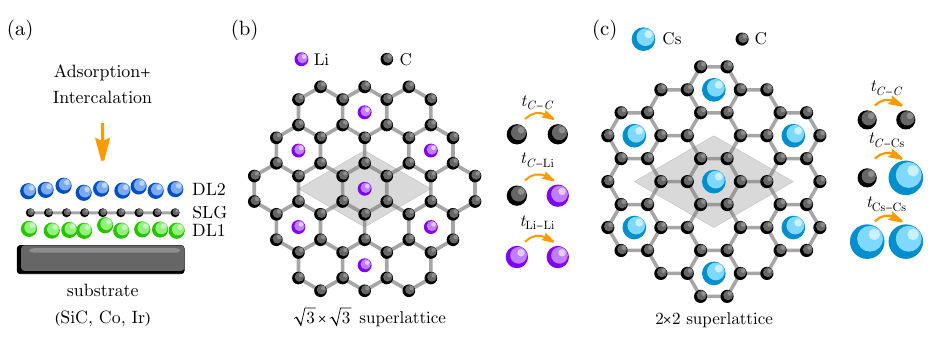}
\caption{(a) Schematic illustration of SLG heavily doped by intercalation and adsorption from side view. A first dopant layer (DL1, green atoms) is intercalated between SLG and a substrate (black rectangle). Additional doping can be introduced by a second dopant layer (DL2, blue atoms) adsorbed on top of the SLG. Dopants on DL1 and DL2 might not necessarily be of the same specie. In some cases, such as for Li and Cs doping, dopants might order periodically and change the lattice symmetry of the SLG, leading to folded $\pi^*$ bands and other features observed in ARPES measurements \cite{Bao2022Coexistence,Ehlen2020Origin}. Here we model them by using superlattice structures as shown in (b, c).
(b) $\sqrt{3}\times\sqrt{3}$ superlattice model for Li-doped SLG. The TB model considers different kinds of hopping integrals: between carbon atoms ($t_{C-C}$), between dopant atoms ($t_{Li-Li}$) and between a carbon and a dopant ($t_{C-Li}$). (c) $2\times2$ superlattice model for Cs-doped SLG. Hopping integrals depicted are defined similarly to those in (b).
\label{Fig: Lattices}}
\end{figure*}

Close to charge neutrality, the electronic properties of graphene are well described by the conventional tight binding (TB) model for electrons in a honeycomb lattice within a first-nearest-neighbors (1NN) approximation. However, angle-resolved photoemission spectroscopy (ARPES) of highly doped graphene reveals that its $\pi$ bands
are strongly renormalized \cite{McChesney2010Extended,Link2019Introducing,Rosenzweig2020Overdoping,Ehlen2020Origin,Bao2022Coexistence,Jugovac2022Clarifying}. The $\pi^*$ (conduction) band is significantly flattened near the M points, leading to a high-order van Hove (HOVH) singularity \cite{Classen2025High}, and the
energy difference between the VH points and Dirac points is $E_{VH}-E_D\approx1.5$ eV, half of what is expected from the 1NN model. It has become evident that this is a robust property of highly doped graphene, with $E_{VH}-E_D$ staying roughly the same across samples, independent of the specie of dopant employed \cite{Rosenzweig2020Overdoping,Ehlen2020Origin,Jugovac2022Clarifying}.
These renormalizations have been attributed to electronic correlations, being reminiscent of signatures in the cuprates and other highly correlated materials \cite{Link2019Introducing}. 
Importantly, the formation of a HOVH singularities is expected to qualitatively modify the many-body phase diagram of highly-doped graphene, affecting the competition between phases \cite{Classen2020Competing}. 

Because doping by intercalation and adsorption involves covering graphene with layers of dopant atoms, its electronic structure may be further changed. Aside from transferring charge to the graphene layer, the dopant atoms can order periodically and form dispersive bands. Most typically, alkaline metals are employed, which tend to form 
$\sqrt{3}\times\sqrt{3}$ and $2\times2$ orders with respect to the graphene lattice (see Fig.~\ref{Fig: Lattices}).  In some cases, the dopant bands might cross $E_F$ and hybridize with the carbon $\pi^*$ states. Additionally, the  order of the dopants can act as a superlattice potential, folding of the graphene bands. Such effects have been recently observed in graphene heavily doped by Li \cite{Bao2021Experimental,Bao2022Coexistence} and Cs \cite{Ehlen2020Origin} which form $\sqrt{3}\times\sqrt{3}$ and $2\times2$ orders, respectively. 

The resulting band structure of these graphene superlattices might lead to nontrivial electronic behavior. 
In some cases, their electronic bands are similar to that of graphite-intercalated compounds \cite{Dresselhaus2002Intercalation}, some of which are established phonon-driven superconductors. Because of this similarity, superconductivity is also expected to occur in some of the doped monolayers \cite{Savini2010first,Profeta2012Phonon,Chapman2016Superconductivity,Ludbrook2015Evidence}.
Moreover, graphene superlattices with $\sqrt{3}\times\sqrt{3}$ order are expected to exhibit topological electronic transport \cite{Wu2016Topological,Freeney2020Edge,Garcia2024Atomically}. Therefore, models that adequately capture the band structure of doped graphene superlattices
might be helpful in studying these doping-induced effects.

Recently, we have studied the superconducting properties of highly-doped graphene~\cite{Herrera2024Topological}.
This work is a follow-up where we study in-depth the technical details of the tight-binding models and electronic properties of these systems, specifically those where the doping agents form an ordered crystalline structure.
Furthermore, we employ the TB models to compute their optical features.
We show that taking into account the periodic ordering of dopants, the renormalization of the $\pi$ bands, and their hybridization with dopant orbitals, one can reproduce the general features seen in ARPES experiments of highly doped graphene \cite{Ehlen2020Origin}.
In Sec II we introduce the TB models, write their explicit matrix forms and report the signatures in their optical conductivity. In Sec. III we present a discussion on the effectiveness and applicability of the models, as well as on possible generalizations.

\section{Effective tight-binding models}

In this section we construct minimal TB models that capture the electronic features discussed above. We focus on effective superlattice structures that include both carbon atoms and dopant atoms ordered with the appropriate symmetry as determined from ARPES and diffraction experiments on intercalation-doped graphene. The models also reproduce the $\pi^*$-band renormalization by including higher-order hopping terms along carbon atoms, which can allow to capture the flattening of the pi band in a phenomenological way.  The parameters of tight binding models are then extracted from DFT calculations. The combined features considered in our models allow us to reproduce the dispersions seen by ARPES experiments in Cs-doped graphene \cite{Ehlen2020Origin}.

\subsection{Description of the models}

\begin{figure}
\includegraphics[width=0.4\textwidth]{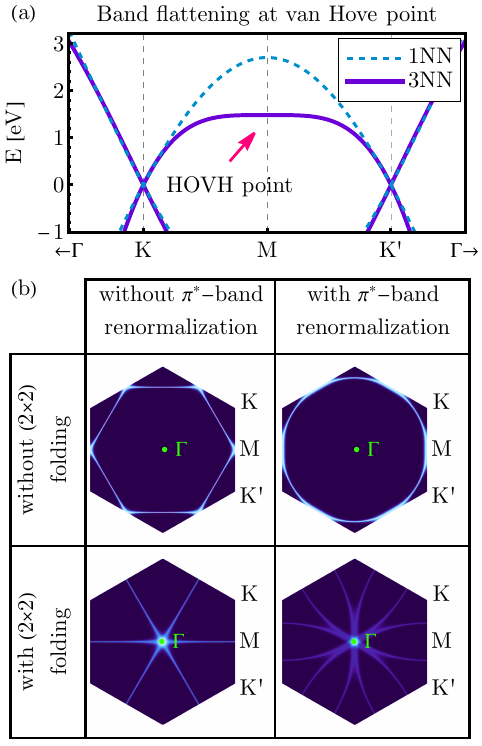}
\caption{Main features in the $\pi^*$-band of SLG that result from heavy doping. (a) Comparison between the band structure for the 1NN and 3NN models. The latter displays bands that are significantly flattened close the the M points, which leads to a HOVH point, as seen in multiple ARPES experiments \cite{McChesney2010Extended,Link2019Introducing,Rosenzweig2020Overdoping}. 
Aside from the band flattening, the lattice symmetry of SLG can change due to dopants ordering periodically. The resulting FSs for both models at van Hove doping considering these effects are shown in (b). 
The FSs on the right panels qualitatively match ARPES data \cite{Ehlen2020Origin,Herrera2024Topological}. Renormalized were bands computed with C-C hoppings $[t,t_2,t_3]=-[4.08,0.93,0.56]$ eV, which approximately satisfy the condition for a HOVH point \cite{Classen2020Competing}.
The folded BZ was scaled to be the same size as the unfolded case.}
\label{Fig: FS}
\end{figure}

Heavy-chemical doping of graphene involves intercalating layers of electron-dopant atoms in between the monolayer and a substrate [see Fig. \ref{Fig: Lattices} (a) for a schematic illustration]. Diffraction experiments indicate that dopants usually form a $2\times2$ and $\sqrt{3}\times\sqrt{3}$ orders with respect to the graphene lattice. This can lead to dopants forming dispersive electronic bands, which can hybridize with the $\pi^*$ band of graphene, opening sizable gaps and overall, significantly modifying the electronic dispersion of SLG.

The effect of dopants on the electronic structure of graphene can be taken into account in effective TB models by considering superlattices as shown in Fig \ref{Fig: Lattices} (b), (c).  Here, we focus on Li-doped and Cs-doped SLG superlattices, which have been seen to form $2\times2$ \cite{Petrovic2013the,Hell2020massive,Ehlen2020Origin} and $\sqrt{3}\times\sqrt{3}$ \cite{Sugawara2011Fabrication,Ludbrook2015Evidence,Bao2021Experimental,Bao2022Coexistence,Ichinokura2022Van,Wu2023Effects} orders with respect to the graphene lattice. Other alkali dopants are expected to order similarly \cite{Dresselhaus2002Intercalation,Csanyi2005,Kanetani2012ca,Huempfner2023Superconductivity}.
Including dopant-dopant and dopant-carbon hoppings in the TB models leads to dispersive dopant bands and their hybridization with $\pi^*$-bands, respectively.
Later we will see that adding these terms is necessary to reproduce the features seen in ARPES experiments.

Following the superlattice structures depicted in Fig. \ref{Fig: Lattices} (b, c), the resulting TB matrices can be written as

\begin{align}\label{eq.H_SLG-Li}
   H_{Li-doped}(\vk)= 
    \begin{bmatrix}
        h_{C}^{(\sqrt{3}\times\sqrt{3})} & h_{C-Li} \\
        h_{C-Li}^\dagger & h_{Li}
    \end{bmatrix} \text{ and}
\end{align}

\begin{align} \label{eq.H_SLG-Cs}
   H_{Cs-doped}(\vk)= 
    \begin{bmatrix}
        h_{C}^{(2\times2)} & h_{C-Cs} \\
        h_{C-Cs}^\dagger & h_{Cs}
    \end{bmatrix},
\end{align}
where $h_C^{(\sqrt{3}\times\sqrt{3})}$ and $h_C^{(2\times 2)}$ describe a $\sqrt{3}\times\sqrt{3}$ and $2\times2$ cell of carbon orbitals in graphene, respectively  and include the 1st, 2nd  and 3rd-NN hoppings ($t_1,t_2,t_3$) (see Appendix). $h_{Li/Cs}$ corresponds to the dopant orbitals, including Li-Li (or Cs-Cs) hoppings which lead to dispersive dopant bands. These are given by
\begin{align}
    h_{Li}&=\epsilon_{Li}
    +2t_{Li2}[\cos(6\vd_2^{Li}\cdot\vk+3\vd_1^{Li}\cdot\vk)\nonumber\\&+
    \cos(6\vd_1^{Li}\cdot\vk+3\vd_2^{Li}\cdot\vk)+
    \cos(3\vd_2^{Li}\cdot\vk-3\vd_1^{Li}\cdot\vk)]\nonumber\\
    &+2t_{Li1}\sum_{i=1}^3\cos(3\vd_i^{Li}\cdot \vk),\\
    h_{Cs}&=\epsilon_{Cs}+2t_{Cs}[\cos(2\vd_1^{Cs}\cdot\vk-2\vd_3^{Cs}\cdot\vk) \nonumber\\
    &+\cos(2\vd_2^{Cs}\cdot\vk-2\vd_3^{Cs}\cdot\vk)+\cos(2\vd_1^{Cs}\cdot\vk-2\vd_2^{Cs}\cdot\vk)],
\end{align}

where $\epsilon_{Li/Cs}$ are the dopant on-site energies and $t_{Li/Cs}$ the 1NN hoppings among dopants. For the G-Li superlattice we have introduced second neighbor hoppings $t_{Li2}$. Here $\vd_i^{Li}$ ($\vd_i^{Cs}$) are the vectors connecting nearest neighbors in the SLG-Li (SLG-Cs) superlattice. For Li-doped SLG $\vd_{1/2}^{Li}=(\mp\sqrt{3},-1)a_c/2$, $\vd_3^{Li}=-\vd_1^{Li}-\vd_2^{Li}$, while for Cs-doped SLG $\vd_{1/2}^{Cs}=(-1,\pm\sqrt{3})a_c/2$, $\vd_3^{Cs}=-\vd_1^{Cs}-\vd_2^{Cs}$ [see Fig. \ref{Fig: Lattices} (b), (c)].
The non-diagonal elements in Eqs. (\ref{eq.H_SLG-Li}-\ref{eq.H_SLG-Cs}) contain the carbon-dopant hoppings $t_{C-Li/Cs}$ leading to hybridization between $\pi^*$ and dopant bands, given by

\begin{align}
    h_{C-Li}=&t_{C-Li}
    \begin{bmatrix}
        \g_1 &
        \g_1^* &
        \g_3 &
        \g_3^* &
        \g_2 &
        \g_2^*
    \end{bmatrix}^\intercal \text{ and}
    \\
     h_{C-Cs}=&t_{C-Cs}
    \begin{bmatrix}
        0 &
        \eta_3 &
        \eta_1^* &
        \eta_2^* &
        \eta_2 &
        \eta_1 &
        \eta_3^* &
        0
    \end{bmatrix}^\intercal,
\end{align}
with $\g_i=\exp(i\vd_i^{Li}\cdot\vk)$ and $\eta_i=\exp(i\vd_i^{Cs}\cdot\vk)$.

\begin{table*}
\caption{\label{tab:table}
Parameters of tight binding models fitted from DFT calculations. Values are shown in units of graphene's 1NN hopping parameter, $t_0=-2.7$ eV.}
\begin{ruledtabular}
\begin{tabular}{cccccccccc}
 Atom & $\epsilon_c$ & $\epsilon_{c}'$ & $\epsilon_{Cs/Li}$ & $t$ & $t_2$ & $t_3$ & $t_{Cs/Li}$ &$ t_{Cs2/Li2}$ & $t_{C-Cs/C-Li}$\\
\hline
Cs& 0.5 & 0.52 & -0.4 & 1.1 & 0.015 & 0.1 & 0.12 & 0 & 0.025\\
Li& 0.695 & 0.81 & -0.38 & 1.28 & 0.08 & 0.2 & 0.135 & -0.04 & 0.05\\
\end{tabular}
\end{ruledtabular}
\label{table2}
\end{table*}

\subsection{General features of the model}
As we have stated before, in addition to the dopant bands and the dopant-carbon hybridization, heavily doped graphene also exhibits a strong renormalization of the $\pi$-bands \cite{McChesney2010Extended,Link2019Introducing,Rosenzweig2020Overdoping,Ehlen2020Origin,Bao2022Coexistence,Jugovac2022Clarifying}. Most importantly, the $\pi^*$ band is significantly flattened near the M points, possibly leading to a HOVHS \cite{Classen2025High}. This appears to be a robust property of highly doped graphene, arising even in the absence of dopant bands near $E_F$, and leading robustly to $E_{VH}-E_D\approx 1.5$ eV (half of what is expected from the 1NN model), independent of the specie of dopant employed. Although this renormalization may have a many-body origin \cite{Link2019Introducing}, it can be captured phenomenologically within a single-particle model by including higher order C-C hoppings in the graphene lattice \cite{McChesney2010Extended,Classen2020Competing}. 
In particular, it was shown that when $t_3=(t_1-2t_2)/4$, the VH points become HOVH points \cite{Classen2020Competing}. 
In Fig. \ref{Fig: FS} (a) we show the renormalized bands of a TB model with up to 3NN hoppings satisfying the condition for a HOVH in comparison to those of the 1NN which show a dispersive band. The former bands were obtained to match previously reported ARPES data for Tb-doped SLG in Ref.~\cite{Herrera2024Topological}, which exhibits renormalization, but not BZ folding since Tb dopants (contrary to Li and Cs dopants) do not exhibit periodic ordering.
For the TB models discussed here, we will take into account the combined effect of renormalization and BZ folding.

Figure \ref{Fig: FS} (b) shows the FSs at van Hove doping that result from the different models with 1NN and 3NN (left and right columns, respectively) and different superstructures (top and bottom rows). 
While the 1NN TB model of graphene exhibits an hexagonal FS (top left), the 3NN model that takes into account the $\pi^*$-band renormalization exhibits a rounded FS (top right). A rounded FS matches well with ARPES experiments on graphene doped by Ca \cite{McChesney2010Extended}, Gd \cite{Link2019Introducing}, Yb \cite{Rosenzweig2020Overdoping} and Tb \cite{Herrera2024Topological}, where high doping is achieved without inducing band folding. Including band folding further modifies the FS (bottom panels). In particular, considering both the $\pi^*$-band renormalization and a $2\times2$ folding leads to a flower-like FS. This shape closely matches recent ARPES experiments on Cs-doped graphene \cite{Ehlen2020Origin}.
Analogously, considering a $\sqrt{3}\times\sqrt{3}$ folding reproduces the FS reported for Li-doped graphene \cite{Sugawara2011Fabrication, Bao2021Experimental,Bao2022Coexistence,Ichinokura2022Van}. Therefore, our effective models are able to qualitatively reproduce the electronic features seen in ARPES experiments.

\begin{figure*}
\includegraphics[width=\textwidth]{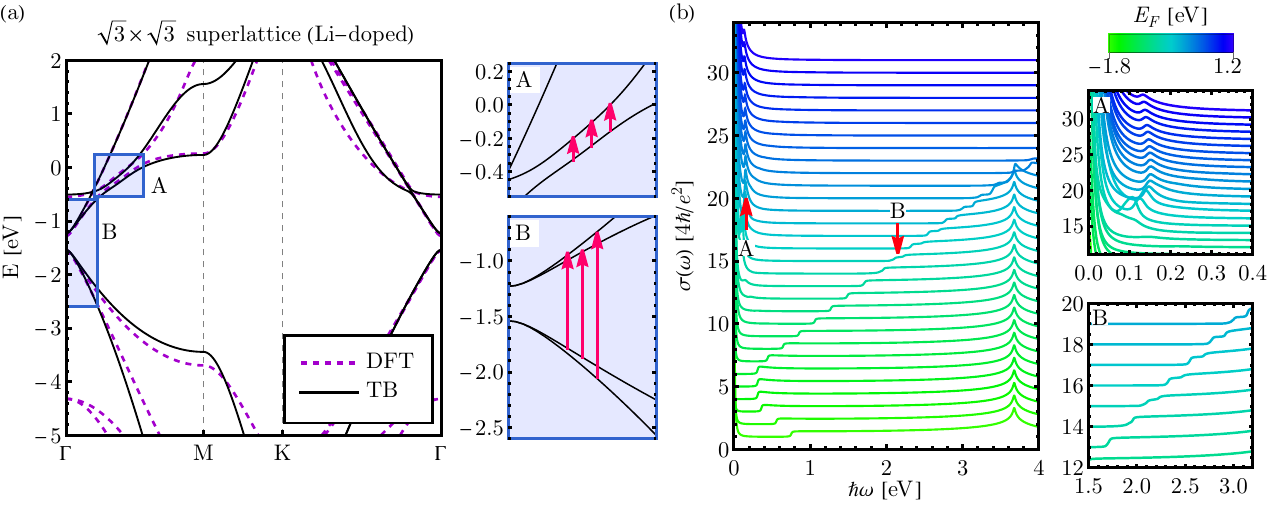}
\caption{Electronic structure and optical signatures of Li-doped SLG. (a) Comparison of electronic bands obtained from DFT calculations and the TB model. The bands exhibit the expected $\sqrt{3}\times\sqrt{3}$ folding induced by the ordering of Li dopants. Due to the folding, the Dirac cones of SLG have been shifted to $\Gamma$, where they hybridize and open a gap, as seen in recent ARPES measurements \cite{Bao2022Coexistence}. The bands enclosed by the blue rectangles (zoom in shown in the right panels), labeled A and B, lead to optical transitions that are expected have characteristic signatures in the optical conductivity. (b) Optical conductivity calculated for the TB model of Li-doped SLG at multiple values of $E_F$ (green-blue scale). Curves are vertically shifted for clarity. Aside from the well-known step-like behavior for SLG, two signatures (marked by red arrows) arise due to the heavy Li-doping. Such signatures are attributed to the corresponding optical transitions shown in panels A and B of (a). Transitions A lead to a low-frequency peak for fillings above $\sim-0.5$ eV. Transitions B lead to a splitting of the step in the optical conductivity of SLG.
\label{Fig: Li}}
\end{figure*}

\subsection{Fitting to first-principles calculations}
Although by just fitting the first, second and third NN hoppings of our TB model we can accurately describe phenomenologically the band structure of high-doped graphene when the doping atoms do not have an ordered structure, in the previous section we have developed an atomistic TB model to obtain the band structure of Li-doped and Cs-doped SLG.
In order to fit the parameters our TB models, we have carried out Density Functional Theory calculations  using a numerical atomic orbitals approach to DFT~\cite{kohsha1965,HohKoh1964} as implemented in the \textsc{Siesta} code~\cite{ArtAng2008,SolArt2002,siesta-2020}.
Core electrons were replaced by norm-conserving scalar relativistic pseudopotentials~\cite{vansetten2018dojo,garcia2018psml} and we use the generalized gradient approximation, specifically the exchange-correlation functional developed by Perdew \emph{et al.}~\cite{PBE96}.
To correctly describe the distance between graphene and the dopant agent, we use the Grimme approximation~\cite{grimme2006semiempirical}. 
We have used a split-valence double-$\zeta $ basis set for all the atoms. For C and Cs we include polarization functions~\cite{arsan99}. 
The energy cutoff of the real space integration mesh was set to 1000 Ry and the brillouin zone was sampled using the Monkhorst-Pack scheme~\cite{MonPac76} with grids of (11$\times$11$\times$1). 
We obtain a lattice constant for graphene of 2.49 \AA.
After building the superstructures for Cs and Li, we relaxed the structure, only allowing for the alkali atoms to move in the $z$-direction with a threshold of 0.01 eV/\AA. We report the TB parameters in Table \ref{tab:table}. 

\begin{figure*}[t]
\includegraphics[width=\textwidth]{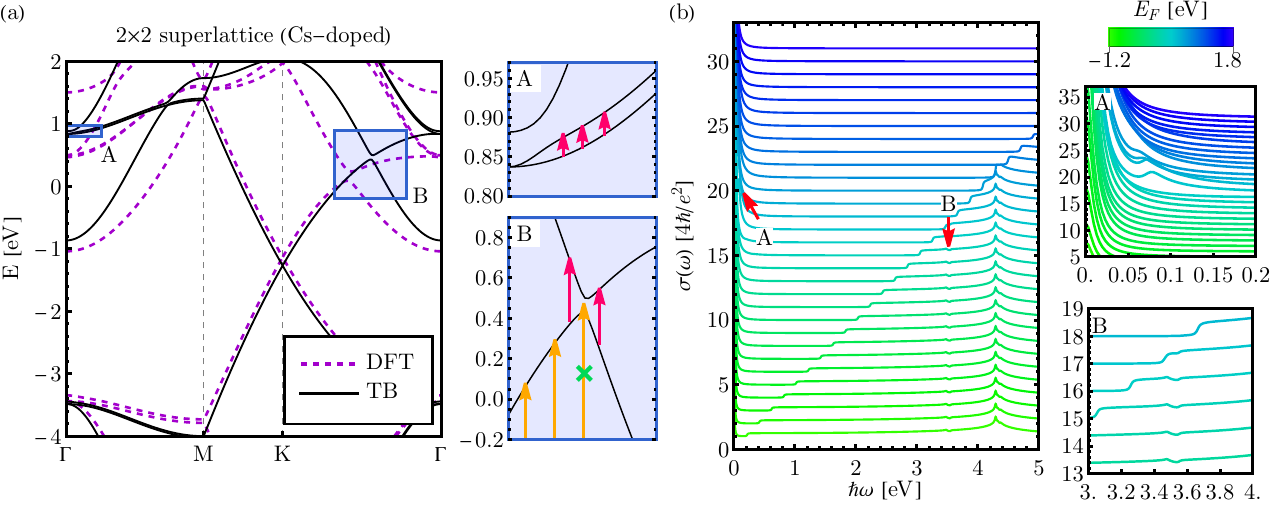}
\caption{Electronic structure and optical signatures of Cs-doped SLG. (a) Comparison of electronic bands obtained from DFT calculations and the TB model. The bands exhibit the expected $2\times2$ folding induced by the ordering of Cs dopants. Due to hybridization between the $\pi^*$-band of SLG a Cs band, a gap is opened between K and $\Gamma$, as seen in recent ARPES measurements \cite{Ehlen2020Origin}. The bands enclosed by the blue rectangles (zoom in shown in the right panels), labeled A and B, lead to optical transitions that are expected have characteristic signatures in the optical conductivity. (b) Optical conductivity calculated for the TB model of Cs-doped SLG at multiple values of $E_F$ (green-blue scale). Curves are vertically shifted for clarity. Aside from the well-known step-like behavior for SLG, two signatures (marked by red arrows) arise due to the heavy Cs-doping. Such signatures are attributed to the corresponding optical transitions shown in panels A and B of (a). Transitions marked by red arrows in panels A and B of (a) lead to the low-frequency peak for fillings above $\sim0.8$ eV.
Transitions marked by orange arrows in panel B lead to the step shape at higher frequencies. There is a dip in the optical conductivity due to these transitions being suppressed by the hybridization gap (crossed orange arrow in panel B).
\label{Fig: Cs}}
\end{figure*}

In Fig. \ref{Fig: Li} (a) we show the fitting of the TB bands to those obtained with DFT for Li-doped graphene. Figure \ref{Fig: Cs} (a) shows similar results for Cs-doped graphene. A good match between DFT and TB bands is obtained.

\section{Optical conductivity}

In the following we discuss the optical signatures of the models presented above. In particular, we compute the optical conductivity \cite{Moon2013optical}
\begin{equation}
    \sigma_{xx}(\omega)=\frac{e^2\hbar}{iS}\sum_{\alpha,\beta}\frac{f(\varepsilon_\alpha)-f(\varepsilon_\beta)}{\varepsilon_\alpha-\varepsilon_\beta}\frac{|\langle\alpha|v_x|\beta\rangle|^2}{\varepsilon_\alpha-\varepsilon_\beta+\hbar\omega+i\eta_0},
\end{equation}
where $\varepsilon_\alpha, \varepsilon_\beta$ and $|\alpha\rangle, |\beta\rangle$ are the energies and eigenstates obtained from the TB hamiltonians with the summation running over all states, $S$ is the area of the system, $f(\varepsilon)=[1+\exp^{(\varepsilon-\mu)/k_B T}]^{-1}$, $\mu$ the chemical potential, $T$ the temperature, and $\eta_0$ a phenomenological broadening. 

The optical conductivity computed with the Li-doped graphene model is shown in Fig. \ref{Fig: Li} (b). It exhibits the same characteristic step-like shape of non-doped SLG, $\sigma(\omega)\sim \Theta[\hbar\omega-2\mu]$, but with some new features arising from transitions between the folded bands. In particular, transitions between close high-energy bands lead to a small low frequency peak (see panel A) at sufficiently high doping. 
At intermediate doping, a double-step shape arises (see panel B) due to the bands at low energy consisting of two gapped Dirac cones at $\Gamma$ with slightly different Fermi velocities, $v_f\pm\delta v$ (with $v_f$ the fermi velocity of nondoped graphene). The different Fermi velocities arise as a consequence of Kekul\'e patterns that form naturally in Li-doped graphene \cite{Gamayun2018valley,Bao2021Experimental}. As a consequence, the step-like conductivity splits into two steps,
separated by a frequency $\Delta\omega\sim 4\mu \delta v/v_f$ \cite{Herrera2020dynamic} which can serve as a measurement of the strength of the Kekul\'e pattern.

Similar results are shown in Fig. \ref{Fig: Cs} (b)
for the conductivity of Cs-doped graphene. Feature A also shows a low energy peak that arises from transitions between high energy bands, although with a different doping dependence than that of Li-doped graphene [Fig. \ref{Fig: Li} (b)]. Contrary to the Li-doped case, feature B is a notable dip in $\sigma(\omega)$ at intermediate and low dopings. This dip arises due to the gap that opens where the $\pi^*$ and dopant bands hybridize [see bottom right panel in Fig. \ref{Fig: Cs} (a)]. The gap is clearly seen in the experimental ARPES bands of Ref. \cite{Ehlen2020Origin}, with an apparent size of $\approx 0.2$ eV.

Therefore, the size of the dip in $\sigma(\omega)$ might provide an optical measure for the strength of carbon-dopant hybridization in doped graphene. This can help characterize samples with $2\times2$ orders, as is the case for Cs doping \cite{Petrovic2013the,Hell2020massive,Ehlen2020Origin} and K doping \cite{Huempfner2023Superconductivity}. In the case of our TB model, the size of the dip is directly related to the C-Cs hopping $t_{C-Cs}$.

\section{Discussion and conclusions}

We have developed effective TB models for highly doped graphene superlattices, where dopants order periodically and modify the lattice symmetry of the monolayer. The TB parameters were fitted to DFT calculations, and the resulting models yield distinctive optical conductivity signatures that can be used to probe the superlattice symmetry and the strength of dopant–carbon hybridization in experiments.

Recent studies of Cs-doped \cite{Ehlen2020Origin,Hell2020massive} and Li-doped graphene~\cite{Sugawara2011Fabrication,Ludbrook2015Evidence,Bao2021Experimental,Bao2022Coexistence,Ichinokura2022Van} have shown that the doped $\pi$ bands deviate markedly from those of the standard honeycomb TB model. Because Cs and Li dopants arrange into $2\times2$ and $\sqrt{3}\times\sqrt{3}$ orders, respectively, they may form dispersive bands that hybridize with carbon orbitals. Our models account for these effects by introducing effective superlattice structures together with dopant–dopant and dopant–carbon hoppings. In addition, experiments reveal a strong flattening of the $\pi^*$ band near the VH point, leading to a HOVHS. This renormalization is incorporated phenomenologically in our models through higher-order carbon–carbon hoppings up to third nearest neighbors \cite{McChesney2010Extended,Classen2020Competing}. By including both superlattice folding and band renormalization, our models reproduce the dispersions observed in multiple ARPES experiments on highly doped graphene~\cite{Ehlen2020Origin,Rosenzweig2020Overdoping,Bao2021Experimental,Bao2022Coexistence}.

The resulting dopant-dependent band structures are expected to strongly influence the many-body phase diagram of VH-doped graphene~\cite{Black2007Resonating,Gonzalez2008Kohn,Nandkishore2012Chiral,Kiesel2012Competing,BlackSchaffer2014Chiral}, particularly through the emergence of higher-order singularities \cite{Classen2025High} and the distinct Fermi surface geometries arising from different superlattice symmetries. Whereas the $\sqrt{3}\times\sqrt{3}$ order preserves the VH points at the M points, the $2\times2$ order folds them to $\Gamma$, suppressing scattering processes with large momentum transfer \cite{Ojajarvi2024pairing} and altering the balance between competing correlated and superconducting phases~\cite{Classen2020Competing,Ojajarvi2024pairing,Herrera2024Topological}. On the single-particle side, superlattice symmetry can also modify transport at lower doping levels, as periodic $\sqrt{3}\times\sqrt{3}$ modulations are predicted to generate topological electronic responses~\cite{Wu2016Topological,Freeney2020Edge,Garcia2024Atomically}.

In practice, highly doped graphene samples might exhibit more complex structures~\cite{McChesney2010Extended,Hell2020massive,Rosenzweig2020Overdoping,Ichinokura2022Van,Zaarour2023Flat}. In addition to an intercalated dopant layer between graphene and the substrate, a different atomic species may be adsorbed on top to further increase doping~\cite{McChesney2010Extended,Rosenzweig2020Overdoping,Ehlen2020Origin}. Moreover, band folding is not always observed, even when dopants order periodically~\cite{McChesney2010Extended,Jugovac2022Clarifying}, likely due to sample-dependent distances between graphene and dopant layers \cite{Profeta2012Phonon}. By contrast, the flattening of the $\pi^*$ band appears as a robust feature of highly doped graphene~\cite{McChesney2010Extended,Rosenzweig2020Overdoping,Ehlen2020Origin,Bao2021Experimental,Bao2022Coexistence}, although it may be partially suppressed by the choice of substrate~\cite{Jugovac2022Clarifying}, possibly due to enhanced Coulomb screening. Nonetheless, our models provide a minimal framework for understanding the electronic structure of highly doped graphene. They capture the essential effects of dopant ordering and band renormalization, and they can be applied to interpret optical signatures, transport properties, and the competition between correlated phases in this class of graphene superlattices.

\section*{Acknowledgments}
We thank Federico Escudero, Zhen Zhan and Danna Liu for useful discussions. IMDEA Nanociencia acknowledges support from the ‘Severo Ochoa’ Programme for Centres of Excellence in R\&D (CEX2020-001039-S/AEI/10.13039/501100011033). 
We acknowledge support from NOVMOMAT, project PID2022-142162NB-I00 funded by MICIU/AEI/10.13039/501100011033 and by FEDER, UE as well as financial support through the (MAD2D-CM)-MRR MATERIALES AVANZADOS-IMDEA-NC.
J.A. S.-G. has received financial support through the ``Ram\'on y Cajal'' Fellowship program, grant RYC2023-044383-I financed by MICIU/AEI/10.13039/501100011033 and FSE+.
G.P.-M. is supported by Comunidad de Madrid through the PIPF2022 programme (grant number PIPF-2022TEC-26326). The Flatiron Institute is a
division of the Simons Foundation.

\onecolumngrid
\appendix
\section{}
The Hamiltonians containing the hoppings among carbon orbitals in the different superlattice structures introduced in Eqs. (\ref{eq.H_SLG-Li}-\ref{eq.H_SLG-Cs}) are given by,
\begin{equation}
h_{C}^{(\sqrt{3}\times\sqrt{3})}=
    \begin{bmatrix}
        \epsilon_{c} & t \g_1^*+t_3 \g_1^2 & t_2g^*  & t\g_2^*+t_3 \g_2^2 & t_2 g  &t \g_3^* +t_3 \g_3^2 \\
        t\g_1+t_3\g_1^{*2} & \epsilon_{c}' & t\g_2+t_3\g_2^{*2} & t_2g  &t\g_3+t_3\g_3^{*2} & t_2 g^*  \\
        t_2g  & t\g_2^*+t_3\g_2^2 & \epsilon_{c} & t\g_3^*+t_3\g_3^2 & t_2g^*  & t\g_1^*+t_3\g_1^2 \\
        t\g_2+t_3\g_2^{*2} & t_2g^*  & t\g_3+t_3\g_3^{*2} & \epsilon_{c}' & t\g_1+t_3\g_1^{*2} & t_2g  \\
        t_2g^*  & t\g_3^*+t_3\g_3^2 & t_2g  &t\g_1^*+t_3\g_1^2 & \epsilon_{c} & t\g_2^*+t_3\g_2^2 \\
        t\g_3+t_3\g_3^{*2} & t_2g  & t\g_1+t_3\g_1^{*2} & t_2g^*  & t\g_2+t_3\g_2^{*2} & \epsilon_{c}'
    \end{bmatrix}
\end{equation}

\begin{equation}
h_{C}^{(2\times2)}=
    \begin{bmatrix}
        \epsilon_c & t \eta_3 & 2t_2c_{23} & 2t_2c_{31} & t \eta_2 & t \eta_1 & 2t_2c_{12} & t_3 h_3^*\\
        t \eta_3^* & \epsilon_c' & t \eta_2^* & t \eta_1^* & 2t_2c_{23} & 2t_2c_{13} & t_3 h_3 & 2t_2c_{12} \\
        2t_2c_{23} & t \eta_2 & \epsilon_c' &2t_2c_{21} & t \eta_3 & t_3 h_3^* & 2t_2c_{13} & t \eta_1 \\
        2t_2c_{13} & t \eta_1 & 2t_2c_{12} & \epsilon_c' & t_3 h_3^* & t \eta_3 & 2t_2c_{23} & t \eta_2 \\
        t \eta_2^* & 2t_2c_{23} & t \eta_3^* & t_3 h_3 & \epsilon_c' & 2t_2c_{12} & t \eta_1^* & 2t_2c_{13} \\
        t \eta_1^* & 2t_2c_{13} & t_3 h_3 & t \eta_3^* & 2t_2c_{12} & \epsilon_c' & t \eta_2^* & 2t_2c_{23} \\
        2t_2c_{12} & t_3 h_3^* & 2t_2c_{13} & 2t_2c_{23} & t \eta_1 & t \eta_2 & \epsilon_c' & t \eta_3 \\
        t_3 h_3 & 2t_2c_{12} & t \eta_1^* & t \eta_2^* & 2t_2c_{13} & 2t_2c_{23} & t \eta_3^* & \epsilon_c
      
    \end{bmatrix}
\end{equation}
here $g=\g_3\g_1^*+\g_1\g_2^*+\g_2\g_3^*$, $h_3=\g_1^*\g_2^*\g_3+\g_1^*\g_2\g_3^*+\g_1\g_2^*\g_3^*$ and $c_{ij}=\cos [(\vd_i^{Cs}-\vd_j^{Cs})\cdot\vk]$.
Second and third NN hoppings, ($t_2$, $t_3$), allow to introduce the flattening and HOVH in the $\pi^*$-band. The basis consists of the carbon orbitals located within the unit cells shown in Fig. \ref{Fig: Lattices} (marked in gray). Ordering of orbitals goes from left to right, top to bottom. Two values of on-site carbon energies $\epsilon_c$ and $\epsilon_c'$, are included to account for the higher electron density at C atoms surrounding dopants, which leads to the band gap at the $\Gamma$ seen in ARPES experiments \cite{Bao2021Experimental} (alternatively, this effect can be captured by considering two different hopping energies). We also note that, although both the Cs- and Li-doped monolayers exhibit some degree of band flattening and require up to 3NN C-C hoppings to be fitted (see Table \ref{tab:table}), the condition for a HOVH \cite{Classen2020Competing} is not met. It has been argued that due to its many-body origin, conventional DFT calculations cannot fully reproduce the HOVH \cite{Link2019Introducing}.

\twocolumngrid

\bibliographystyle{apsrev4-2}
\bibliography{References_exported,References_added}

\begin{thebibliography}{58}%
\makeatletter
\providecommand \@ifxundefined [1]{%
 \@ifx{#1\undefined}
}%
\providecommand \@ifnum [1]{%
 \ifnum #1\expandafter \@firstoftwo
 \else \expandafter \@secondoftwo
 \fi
}%
\providecommand \@ifx [1]{%
 \ifx #1\expandafter \@firstoftwo
 \else \expandafter \@secondoftwo
 \fi
}%
\providecommand \natexlab [1]{#1}%
\providecommand \enquote  [1]{``#1''}%
\providecommand \bibnamefont  [1]{#1}%
\providecommand \bibfnamefont [1]{#1}%
\providecommand \citenamefont [1]{#1}%
\providecommand \href@noop [0]{\@secondoftwo}%
\providecommand \href [0]{\begingroup \@sanitize@url \@href}%
\providecommand \@href[1]{\@@startlink{#1}\@@href}%
\providecommand \@@href[1]{\endgroup#1\@@endlink}%
\providecommand \@sanitize@url [0]{\catcode `\\12\catcode `\$12\catcode `\&12\catcode `\#12\catcode `\^12\catcode `\_12\catcode `\%12\relax}%
\providecommand \@@startlink[1]{}%
\providecommand \@@endlink[0]{}%
\providecommand \url  [0]{\begingroup\@sanitize@url \@url }%
\providecommand \@url [1]{\endgroup\@href {#1}{\urlprefix }}%
\providecommand \urlprefix  [0]{URL }%
\providecommand \Eprint [0]{\href }%
\providecommand \doibase [0]{https://doi.org/}%
\providecommand \selectlanguage [0]{\@gobble}%
\providecommand \bibinfo  [0]{\@secondoftwo}%
\providecommand \bibfield  [0]{\@secondoftwo}%
\providecommand \translation [1]{[#1]}%
\providecommand \BibitemOpen [0]{}%
\providecommand \bibitemStop [0]{}%
\providecommand \bibitemNoStop [0]{.\EOS\space}%
\providecommand \EOS [0]{\spacefactor3000\relax}%
\providecommand \BibitemShut  [1]{\csname bibitem#1\endcsname}%
\let\auto@bib@innerbib\@empty
\bibitem [{\citenamefont {Cao}\ \emph {et~al.}(2018{\natexlab{a}})\citenamefont {Cao}, \citenamefont {Fatemi}, \citenamefont {Demir}, \citenamefont {Fang}, \citenamefont {Tomarken}, \citenamefont {Luo}, \citenamefont {Sanchez-Yamagishi}, \citenamefont {Watanabe}, \citenamefont {Taniguchi}, \citenamefont {Kaxiras}, \citenamefont {Ashoori},\ and\ \citenamefont {Jarillo-Herrero}}]{Cao2018Correlated}%
  \BibitemOpen
  \bibfield  {author} {\bibinfo {author} {\bibfnamefont {Y.}~\bibnamefont {Cao}}, \bibinfo {author} {\bibfnamefont {V.}~\bibnamefont {Fatemi}}, \bibinfo {author} {\bibfnamefont {A.}~\bibnamefont {Demir}}, \bibinfo {author} {\bibfnamefont {S.}~\bibnamefont {Fang}}, \bibinfo {author} {\bibfnamefont {S.~L.}\ \bibnamefont {Tomarken}}, \bibinfo {author} {\bibfnamefont {J.~Y.}\ \bibnamefont {Luo}}, \bibinfo {author} {\bibfnamefont {J.~D.}\ \bibnamefont {Sanchez-Yamagishi}}, \bibinfo {author} {\bibfnamefont {K.}~\bibnamefont {Watanabe}}, \bibinfo {author} {\bibfnamefont {T.}~\bibnamefont {Taniguchi}}, \bibinfo {author} {\bibfnamefont {E.}~\bibnamefont {Kaxiras}}, \bibinfo {author} {\bibfnamefont {R.~C.}\ \bibnamefont {Ashoori}},\ and\ \bibinfo {author} {\bibfnamefont {P.}~\bibnamefont {Jarillo-Herrero}},\ }\href {https://doi.org/10.1038/nature26154} {\bibfield  {journal} {\bibinfo  {journal} {Nature}\ }\textbf {\bibinfo {volume} {556}},\ \bibinfo {pages} {80} (\bibinfo {year} {2018}{\natexlab{a}})}\BibitemShut
  {NoStop}%
\bibitem [{\citenamefont {Cao}\ \emph {et~al.}(2018{\natexlab{b}})\citenamefont {Cao}, \citenamefont {Fatemi}, \citenamefont {Fang}, \citenamefont {Watanabe}, \citenamefont {Taniguchi}, \citenamefont {Kaxiras},\ and\ \citenamefont {Jarillo-Herrero}}]{Cao2018Unconventional}%
  \BibitemOpen
  \bibfield  {author} {\bibinfo {author} {\bibfnamefont {Y.}~\bibnamefont {Cao}}, \bibinfo {author} {\bibfnamefont {V.}~\bibnamefont {Fatemi}}, \bibinfo {author} {\bibfnamefont {S.}~\bibnamefont {Fang}}, \bibinfo {author} {\bibfnamefont {K.}~\bibnamefont {Watanabe}}, \bibinfo {author} {\bibfnamefont {T.}~\bibnamefont {Taniguchi}}, \bibinfo {author} {\bibfnamefont {E.}~\bibnamefont {Kaxiras}},\ and\ \bibinfo {author} {\bibfnamefont {P.}~\bibnamefont {Jarillo-Herrero}},\ }\href {https://doi.org/10.1038/nature26160} {\bibfield  {journal} {\bibinfo  {journal} {Nature}\ }\textbf {\bibinfo {volume} {556}},\ \bibinfo {pages} {43} (\bibinfo {year} {2018}{\natexlab{b}})}\BibitemShut {NoStop}%
\bibitem [{\citenamefont {Lu}\ \emph {et~al.}(2019)\citenamefont {Lu}, \citenamefont {Stepanov}, \citenamefont {Yang}, \citenamefont {Xie}, \citenamefont {Aamir}, \citenamefont {Das}, \citenamefont {Urgell}, \citenamefont {Watanabe}, \citenamefont {Taniguchi}, \citenamefont {Zhang}, \citenamefont {Bachtold}, \citenamefont {MacDonald},\ and\ \citenamefont {Efetov}}]{Lu2019Superconductors}%
  \BibitemOpen
  \bibfield  {author} {\bibinfo {author} {\bibfnamefont {X.}~\bibnamefont {Lu}}, \bibinfo {author} {\bibfnamefont {P.}~\bibnamefont {Stepanov}}, \bibinfo {author} {\bibfnamefont {W.}~\bibnamefont {Yang}}, \bibinfo {author} {\bibfnamefont {M.}~\bibnamefont {Xie}}, \bibinfo {author} {\bibfnamefont {M.~A.}\ \bibnamefont {Aamir}}, \bibinfo {author} {\bibfnamefont {I.}~\bibnamefont {Das}}, \bibinfo {author} {\bibfnamefont {C.}~\bibnamefont {Urgell}}, \bibinfo {author} {\bibfnamefont {K.}~\bibnamefont {Watanabe}}, \bibinfo {author} {\bibfnamefont {T.}~\bibnamefont {Taniguchi}}, \bibinfo {author} {\bibfnamefont {G.}~\bibnamefont {Zhang}}, \bibinfo {author} {\bibfnamefont {A.}~\bibnamefont {Bachtold}}, \bibinfo {author} {\bibfnamefont {A.~H.}\ \bibnamefont {MacDonald}},\ and\ \bibinfo {author} {\bibfnamefont {D.~K.}\ \bibnamefont {Efetov}},\ }\href {https://doi.org/10.1038/s41586-019-1695-0} {\bibfield  {journal} {\bibinfo  {journal} {Nature}\ }\textbf {\bibinfo {volume} {574}},\ \bibinfo {pages} {653} (\bibinfo
  {year} {2019})}\BibitemShut {NoStop}%
\bibitem [{\citenamefont {Mao}\ \emph {et~al.}(2020)\citenamefont {Mao}, \citenamefont {Milovanovi{\'{c}}}, \citenamefont {Andelkovi{\'{c}}}, \citenamefont {Lai}, \citenamefont {Cao}, \citenamefont {Watanabe}, \citenamefont {Taniguchi}, \citenamefont {Covaci}, \citenamefont {Peeters}, \citenamefont {Geim}, \citenamefont {Jiang},\ and\ \citenamefont {Andrei}}]{Mao2020Evidence}%
  \BibitemOpen
  \bibfield  {author} {\bibinfo {author} {\bibfnamefont {J.}~\bibnamefont {Mao}}, \bibinfo {author} {\bibfnamefont {S.~P.}\ \bibnamefont {Milovanovi{\'{c}}}}, \bibinfo {author} {\bibfnamefont {M.}~\bibnamefont {Andelkovi{\'{c}}}}, \bibinfo {author} {\bibfnamefont {X.}~\bibnamefont {Lai}}, \bibinfo {author} {\bibfnamefont {Y.}~\bibnamefont {Cao}}, \bibinfo {author} {\bibfnamefont {K.}~\bibnamefont {Watanabe}}, \bibinfo {author} {\bibfnamefont {T.}~\bibnamefont {Taniguchi}}, \bibinfo {author} {\bibfnamefont {L.}~\bibnamefont {Covaci}}, \bibinfo {author} {\bibfnamefont {F.~M.}\ \bibnamefont {Peeters}}, \bibinfo {author} {\bibfnamefont {A.~K.}\ \bibnamefont {Geim}}, \bibinfo {author} {\bibfnamefont {Y.}~\bibnamefont {Jiang}},\ and\ \bibinfo {author} {\bibfnamefont {E.~Y.}\ \bibnamefont {Andrei}},\ }\href {https://doi.org/10.1038/s41586-020-2567-3} {\bibfield  {journal} {\bibinfo  {journal} {Nature}\ }\textbf {\bibinfo {volume} {584}},\ \bibinfo {pages} {215} (\bibinfo {year} {2020})}\BibitemShut {NoStop}%
\bibitem [{\citenamefont {Zhou}\ \emph {et~al.}(2021)\citenamefont {Zhou}, \citenamefont {Xie}, \citenamefont {Taniguchi}, \citenamefont {Watanabe},\ and\ \citenamefont {Young}}]{Zhou2021SuperRTG}%
  \BibitemOpen
  \bibfield  {author} {\bibinfo {author} {\bibfnamefont {H.}~\bibnamefont {Zhou}}, \bibinfo {author} {\bibfnamefont {T.}~\bibnamefont {Xie}}, \bibinfo {author} {\bibfnamefont {T.}~\bibnamefont {Taniguchi}}, \bibinfo {author} {\bibfnamefont {K.}~\bibnamefont {Watanabe}},\ and\ \bibinfo {author} {\bibfnamefont {A.~F.}\ \bibnamefont {Young}},\ }\href {https://doi.org/10.1038/s41586-021-03926-0} {\bibfield  {journal} {\bibinfo  {journal} {Nature}\ }\textbf {\bibinfo {volume} {598}},\ \bibinfo {pages} {434} (\bibinfo {year} {2021})}\BibitemShut {NoStop}%
\bibitem [{\citenamefont {Zhou}\ \emph {et~al.}(2022)\citenamefont {Zhou}, \citenamefont {Holleis}, \citenamefont {Saito}, \citenamefont {Cohen}, \citenamefont {Huynh}, \citenamefont {Patterson}, \citenamefont {Yang}, \citenamefont {Taniguchi}, \citenamefont {Watanabe},\ and\ \citenamefont {Young}}]{Zhou2022Isospin}%
  \BibitemOpen
  \bibfield  {author} {\bibinfo {author} {\bibfnamefont {H.}~\bibnamefont {Zhou}}, \bibinfo {author} {\bibfnamefont {L.}~\bibnamefont {Holleis}}, \bibinfo {author} {\bibfnamefont {Y.}~\bibnamefont {Saito}}, \bibinfo {author} {\bibfnamefont {L.}~\bibnamefont {Cohen}}, \bibinfo {author} {\bibfnamefont {W.}~\bibnamefont {Huynh}}, \bibinfo {author} {\bibfnamefont {C.~L.}\ \bibnamefont {Patterson}}, \bibinfo {author} {\bibfnamefont {F.}~\bibnamefont {Yang}}, \bibinfo {author} {\bibfnamefont {T.}~\bibnamefont {Taniguchi}}, \bibinfo {author} {\bibfnamefont {K.}~\bibnamefont {Watanabe}},\ and\ \bibinfo {author} {\bibfnamefont {A.~F.}\ \bibnamefont {Young}},\ }\href {https://doi.org/10.1126/science.abm8386} {\bibfield  {journal} {\bibinfo  {journal} {Science}\ }\textbf {\bibinfo {volume} {375}},\ \bibinfo {pages} {774} (\bibinfo {year} {2022})}\BibitemShut {NoStop}%
\bibitem [{\citenamefont {Lu}\ \emph {et~al.}(2022)\citenamefont {Lu}, \citenamefont {Le}, \citenamefont {Zhang}, \citenamefont {Cook}, \citenamefont {He}, \citenamefont {Zarenia}, \citenamefont {Vaninger}, \citenamefont {Miceli}, \citenamefont {Singh}, \citenamefont {Liu}, \citenamefont {Qin}, \citenamefont {Chiang}, \citenamefont {Chiu}, \citenamefont {Vignale},\ and\ \citenamefont {Bian}}]{Qiangsheng2022Dirac}%
  \BibitemOpen
  \bibfield  {author} {\bibinfo {author} {\bibfnamefont {Q.}~\bibnamefont {Lu}}, \bibinfo {author} {\bibfnamefont {C.}~\bibnamefont {Le}}, \bibinfo {author} {\bibfnamefont {X.}~\bibnamefont {Zhang}}, \bibinfo {author} {\bibfnamefont {J.}~\bibnamefont {Cook}}, \bibinfo {author} {\bibfnamefont {X.}~\bibnamefont {He}}, \bibinfo {author} {\bibfnamefont {M.}~\bibnamefont {Zarenia}}, \bibinfo {author} {\bibfnamefont {M.}~\bibnamefont {Vaninger}}, \bibinfo {author} {\bibfnamefont {P.~F.}\ \bibnamefont {Miceli}}, \bibinfo {author} {\bibfnamefont {D.~J.}\ \bibnamefont {Singh}}, \bibinfo {author} {\bibfnamefont {C.}~\bibnamefont {Liu}}, \bibinfo {author} {\bibfnamefont {H.}~\bibnamefont {Qin}}, \bibinfo {author} {\bibfnamefont {T.-C.}\ \bibnamefont {Chiang}}, \bibinfo {author} {\bibfnamefont {C.-K.}\ \bibnamefont {Chiu}}, \bibinfo {author} {\bibfnamefont {G.}~\bibnamefont {Vignale}},\ and\ \bibinfo {author} {\bibfnamefont {G.}~\bibnamefont {Bian}},\ }\href {https://doi.org/https://doi.org/10.1002/adma.202200625}
  {\bibfield  {journal} {\bibinfo  {journal} {Advanced Materials}\ }\textbf {\bibinfo {volume} {34}},\ \bibinfo {pages} {2200625} (\bibinfo {year} {2022})}\BibitemShut {NoStop}%
\bibitem [{\citenamefont {Black-Schaffer}\ and\ \citenamefont {Doniach}(2007)}]{Black2007Resonating}%
  \BibitemOpen
  \bibfield  {author} {\bibinfo {author} {\bibfnamefont {A.~M.}\ \bibnamefont {Black-Schaffer}}\ and\ \bibinfo {author} {\bibfnamefont {S.}~\bibnamefont {Doniach}},\ }\href {https://doi.org/10.1103/PhysRevB.75.134512} {\bibfield  {journal} {\bibinfo  {journal} {Phys. Rev. B}\ }\textbf {\bibinfo {volume} {75}},\ \bibinfo {pages} {134512} (\bibinfo {year} {2007})}\BibitemShut {NoStop}%
\bibitem [{\citenamefont {Uchoa}\ and\ \citenamefont {Castro~Neto}(2007)}]{Uchoa2007Superconducting}%
  \BibitemOpen
  \bibfield  {author} {\bibinfo {author} {\bibfnamefont {B.}~\bibnamefont {Uchoa}}\ and\ \bibinfo {author} {\bibfnamefont {A.~H.}\ \bibnamefont {Castro~Neto}},\ }\href {https://doi.org/10.1103/PhysRevLett.98.146801} {\bibfield  {journal} {\bibinfo  {journal} {Phys. Rev. Lett.}\ }\textbf {\bibinfo {volume} {98}},\ \bibinfo {pages} {146801} (\bibinfo {year} {2007})}\BibitemShut {NoStop}%
\bibitem [{\citenamefont {Gonz\'alez}(2008)}]{Gonzalez2008Kohn}%
  \BibitemOpen
  \bibfield  {author} {\bibinfo {author} {\bibfnamefont {J.}~\bibnamefont {Gonz\'alez}},\ }\href {https://doi.org/10.1103/PhysRevB.78.205431} {\bibfield  {journal} {\bibinfo  {journal} {Phys. Rev. B}\ }\textbf {\bibinfo {volume} {78}},\ \bibinfo {pages} {205431} (\bibinfo {year} {2008})}\BibitemShut {NoStop}%
\bibitem [{\citenamefont {Nandkishore}\ \emph {et~al.}(2012)\citenamefont {Nandkishore}, \citenamefont {Levitov},\ and\ \citenamefont {Chubukov}}]{Nandkishore2012Chiral}%
  \BibitemOpen
  \bibfield  {author} {\bibinfo {author} {\bibfnamefont {R.}~\bibnamefont {Nandkishore}}, \bibinfo {author} {\bibfnamefont {L.~S.}\ \bibnamefont {Levitov}},\ and\ \bibinfo {author} {\bibfnamefont {A.~V.}\ \bibnamefont {Chubukov}},\ }\href {https://doi.org/10.1038/nphys2208} {\bibfield  {journal} {\bibinfo  {journal} {Nat. Phys.}\ }\textbf {\bibinfo {volume} {8}},\ \bibinfo {pages} {158} (\bibinfo {year} {2012})}\BibitemShut {NoStop}%
\bibitem [{\citenamefont {Kiesel}\ \emph {et~al.}(2012)\citenamefont {Kiesel}, \citenamefont {Platt}, \citenamefont {Hanke}, \citenamefont {Abanin},\ and\ \citenamefont {Thomale}}]{Kiesel2012Competing}%
  \BibitemOpen
  \bibfield  {author} {\bibinfo {author} {\bibfnamefont {M.~L.}\ \bibnamefont {Kiesel}}, \bibinfo {author} {\bibfnamefont {C.}~\bibnamefont {Platt}}, \bibinfo {author} {\bibfnamefont {W.}~\bibnamefont {Hanke}}, \bibinfo {author} {\bibfnamefont {D.~A.}\ \bibnamefont {Abanin}},\ and\ \bibinfo {author} {\bibfnamefont {R.}~\bibnamefont {Thomale}},\ }\href {https://doi.org/10.1103/PhysRevB.86.020507} {\bibfield  {journal} {\bibinfo  {journal} {Phys. Rev. B}\ }\textbf {\bibinfo {volume} {86}},\ \bibinfo {pages} {020507} (\bibinfo {year} {2012})}\BibitemShut {NoStop}%
\bibitem [{\citenamefont {Classen}\ \emph {et~al.}(2020)\citenamefont {Classen}, \citenamefont {Chubukov}, \citenamefont {Honerkamp},\ and\ \citenamefont {Scherer}}]{Classen2020Competing}%
  \BibitemOpen
  \bibfield  {author} {\bibinfo {author} {\bibfnamefont {L.}~\bibnamefont {Classen}}, \bibinfo {author} {\bibfnamefont {A.~V.}\ \bibnamefont {Chubukov}}, \bibinfo {author} {\bibfnamefont {C.}~\bibnamefont {Honerkamp}},\ and\ \bibinfo {author} {\bibfnamefont {M.~M.}\ \bibnamefont {Scherer}},\ }\href {https://doi.org/10.1103/PhysRevB.102.125141} {\bibfield  {journal} {\bibinfo  {journal} {Phys. Rev. B}\ }\textbf {\bibinfo {volume} {102}},\ \bibinfo {pages} {125141} (\bibinfo {year} {2020})}\BibitemShut {NoStop}%
\bibitem [{\citenamefont {McChesney}\ \emph {et~al.}(2010)\citenamefont {McChesney}, \citenamefont {Bostwick}, \citenamefont {Ohta}, \citenamefont {Seyller}, \citenamefont {Horn}, \citenamefont {Gonz\'alez},\ and\ \citenamefont {Rotenberg}}]{McChesney2010Extended}%
  \BibitemOpen
  \bibfield  {author} {\bibinfo {author} {\bibfnamefont {J.~L.}\ \bibnamefont {McChesney}}, \bibinfo {author} {\bibfnamefont {A.}~\bibnamefont {Bostwick}}, \bibinfo {author} {\bibfnamefont {T.}~\bibnamefont {Ohta}}, \bibinfo {author} {\bibfnamefont {T.}~\bibnamefont {Seyller}}, \bibinfo {author} {\bibfnamefont {K.}~\bibnamefont {Horn}}, \bibinfo {author} {\bibfnamefont {J.}~\bibnamefont {Gonz\'alez}},\ and\ \bibinfo {author} {\bibfnamefont {E.}~\bibnamefont {Rotenberg}},\ }\href {https://doi.org/10.1103/PhysRevLett.104.136803} {\bibfield  {journal} {\bibinfo  {journal} {Phys. Rev. Lett.}\ }\textbf {\bibinfo {volume} {104}},\ \bibinfo {pages} {136803} (\bibinfo {year} {2010})}\BibitemShut {NoStop}%
\bibitem [{\citenamefont {Efetov}(2014)}]{Efetov2014Towards}%
  \BibitemOpen
  \bibfield  {author} {\bibinfo {author} {\bibfnamefont {D.~K.}\ \bibnamefont {Efetov}},\ }{\selectlanguage {English}\emph {\bibinfo {title} {Towards inducing superconductivity into graphene}}},\ \href@noop {} {Ph.D. thesis} (\bibinfo {year} {2014})\BibitemShut {NoStop}%
\bibitem [{\citenamefont {Rosenzweig}\ \emph {et~al.}(2020)\citenamefont {Rosenzweig}, \citenamefont {Karakachian}, \citenamefont {Marchenko}, \citenamefont {K\"uster},\ and\ \citenamefont {Starke}}]{Rosenzweig2020Overdoping}%
  \BibitemOpen
  \bibfield  {author} {\bibinfo {author} {\bibfnamefont {{\relax Ph}.}~\bibnamefont {Rosenzweig}}, \bibinfo {author} {\bibfnamefont {H.}~\bibnamefont {Karakachian}}, \bibinfo {author} {\bibfnamefont {D.}~\bibnamefont {Marchenko}}, \bibinfo {author} {\bibfnamefont {K.}~\bibnamefont {K\"uster}},\ and\ \bibinfo {author} {\bibfnamefont {U.}~\bibnamefont {Starke}},\ }\href {https://doi.org/10.1103/PhysRevLett.125.176403} {\bibfield  {journal} {\bibinfo  {journal} {Phys. Rev. Lett.}\ }\textbf {\bibinfo {volume} {125}},\ \bibinfo {pages} {176403} (\bibinfo {year} {2020})}\BibitemShut {NoStop}%
\bibitem [{\citenamefont {Efetov}\ and\ \citenamefont {Kim}(2010)}]{Efetov2010Controlling}%
  \BibitemOpen
  \bibfield  {author} {\bibinfo {author} {\bibfnamefont {D.~K.}\ \bibnamefont {Efetov}}\ and\ \bibinfo {author} {\bibfnamefont {P.}~\bibnamefont {Kim}},\ }\href {https://doi.org/10.1103/PhysRevLett.105.256805} {\bibfield  {journal} {\bibinfo  {journal} {Phys. Rev. Lett.}\ }\textbf {\bibinfo {volume} {105}},\ \bibinfo {pages} {256805} (\bibinfo {year} {2010})}\BibitemShut {NoStop}%
\bibitem [{\citenamefont {Rosenzweig}\ \emph {et~al.}(2019)\citenamefont {Rosenzweig}, \citenamefont {Karakachian}, \citenamefont {Link}, \citenamefont {K\"uster},\ and\ \citenamefont {Starke}}]{Rosenzweig2019Tuning}%
  \BibitemOpen
  \bibfield  {author} {\bibinfo {author} {\bibfnamefont {{\relax Ph}.}~\bibnamefont {Rosenzweig}}, \bibinfo {author} {\bibfnamefont {H.}~\bibnamefont {Karakachian}}, \bibinfo {author} {\bibfnamefont {S.}~\bibnamefont {Link}}, \bibinfo {author} {\bibfnamefont {K.}~\bibnamefont {K\"uster}},\ and\ \bibinfo {author} {\bibfnamefont {U.}~\bibnamefont {Starke}},\ }\href {https://doi.org/10.1103/PhysRevB.100.035445} {\bibfield  {journal} {\bibinfo  {journal} {Phys. Rev. B}\ }\textbf {\bibinfo {volume} {100}},\ \bibinfo {pages} {035445} (\bibinfo {year} {2019})}\BibitemShut {NoStop}%
\bibitem [{\citenamefont {Bao}\ \emph {et~al.}(2022)\citenamefont {Bao}, \citenamefont {Zhang}, \citenamefont {Wu}, \citenamefont {Zhou}, \citenamefont {Li}, \citenamefont {Yu}, \citenamefont {Li}, \citenamefont {Duan},\ and\ \citenamefont {Zhou}}]{Bao2022Coexistence}%
  \BibitemOpen
  \bibfield  {author} {\bibinfo {author} {\bibfnamefont {C.}~\bibnamefont {Bao}}, \bibinfo {author} {\bibfnamefont {H.}~\bibnamefont {Zhang}}, \bibinfo {author} {\bibfnamefont {X.}~\bibnamefont {Wu}}, \bibinfo {author} {\bibfnamefont {S.}~\bibnamefont {Zhou}}, \bibinfo {author} {\bibfnamefont {Q.}~\bibnamefont {Li}}, \bibinfo {author} {\bibfnamefont {P.}~\bibnamefont {Yu}}, \bibinfo {author} {\bibfnamefont {J.}~\bibnamefont {Li}}, \bibinfo {author} {\bibfnamefont {W.}~\bibnamefont {Duan}},\ and\ \bibinfo {author} {\bibfnamefont {S.}~\bibnamefont {Zhou}},\ }\href {https://doi.org/10.1103/PhysRevB.105.L161106} {\bibfield  {journal} {\bibinfo  {journal} {Phys. Rev. B}\ }\textbf {\bibinfo {volume} {105}},\ \bibinfo {pages} {L161106} (\bibinfo {year} {2022})}\BibitemShut {NoStop}%
\bibitem [{\citenamefont {Ehlen}\ \emph {et~al.}(2020)\citenamefont {Ehlen}, \citenamefont {Hell}, \citenamefont {Marini}, \citenamefont {Hasdeo}, \citenamefont {Saito}, \citenamefont {Falke}, \citenamefont {Goerbig}, \citenamefont {Di~Santo}, \citenamefont {Petaccia}, \citenamefont {Profeta},\ and\ \citenamefont {Grüneis}}]{Ehlen2020Origin}%
  \BibitemOpen
  \bibfield  {author} {\bibinfo {author} {\bibfnamefont {N.}~\bibnamefont {Ehlen}}, \bibinfo {author} {\bibfnamefont {M.}~\bibnamefont {Hell}}, \bibinfo {author} {\bibfnamefont {G.}~\bibnamefont {Marini}}, \bibinfo {author} {\bibfnamefont {E.~H.}\ \bibnamefont {Hasdeo}}, \bibinfo {author} {\bibfnamefont {R.}~\bibnamefont {Saito}}, \bibinfo {author} {\bibfnamefont {Y.}~\bibnamefont {Falke}}, \bibinfo {author} {\bibfnamefont {M.~O.}\ \bibnamefont {Goerbig}}, \bibinfo {author} {\bibfnamefont {G.}~\bibnamefont {Di~Santo}}, \bibinfo {author} {\bibfnamefont {L.}~\bibnamefont {Petaccia}}, \bibinfo {author} {\bibfnamefont {G.}~\bibnamefont {Profeta}},\ and\ \bibinfo {author} {\bibfnamefont {A.}~\bibnamefont {Grüneis}},\ }\href {https://doi.org/10.1021/acsnano.9b08622} {\bibfield  {journal} {\bibinfo  {journal} {ACS Nano}\ }\textbf {\bibinfo {volume} {14}},\ \bibinfo {pages} {1055} (\bibinfo {year} {2020})}\BibitemShut {NoStop}%
\bibitem [{\citenamefont {Link}\ \emph {et~al.}(2019)\citenamefont {Link}, \citenamefont {Forti}, \citenamefont {St\"ohr}, \citenamefont {K\"uster}, \citenamefont {R\"osner}, \citenamefont {Hirschmeier}, \citenamefont {Chen}, \citenamefont {Avila}, \citenamefont {Asensio}, \citenamefont {Zakharov}, \citenamefont {Wehling}, \citenamefont {Lichtenstein}, \citenamefont {Katsnelson},\ and\ \citenamefont {Starke}}]{Link2019Introducing}%
  \BibitemOpen
  \bibfield  {author} {\bibinfo {author} {\bibfnamefont {S.}~\bibnamefont {Link}}, \bibinfo {author} {\bibfnamefont {S.}~\bibnamefont {Forti}}, \bibinfo {author} {\bibfnamefont {A.}~\bibnamefont {St\"ohr}}, \bibinfo {author} {\bibfnamefont {K.}~\bibnamefont {K\"uster}}, \bibinfo {author} {\bibfnamefont {M.}~\bibnamefont {R\"osner}}, \bibinfo {author} {\bibfnamefont {D.}~\bibnamefont {Hirschmeier}}, \bibinfo {author} {\bibfnamefont {C.}~\bibnamefont {Chen}}, \bibinfo {author} {\bibfnamefont {J.}~\bibnamefont {Avila}}, \bibinfo {author} {\bibfnamefont {M.~C.}\ \bibnamefont {Asensio}}, \bibinfo {author} {\bibfnamefont {A.~A.}\ \bibnamefont {Zakharov}}, \bibinfo {author} {\bibfnamefont {T.~O.}\ \bibnamefont {Wehling}}, \bibinfo {author} {\bibfnamefont {A.~I.}\ \bibnamefont {Lichtenstein}}, \bibinfo {author} {\bibfnamefont {M.~I.}\ \bibnamefont {Katsnelson}},\ and\ \bibinfo {author} {\bibfnamefont {U.}~\bibnamefont {Starke}},\ }\href {https://doi.org/10.1103/PhysRevB.100.121407} {\bibfield  {journal} {\bibinfo
  {journal} {Phys. Rev. B}\ }\textbf {\bibinfo {volume} {100}},\ \bibinfo {pages} {121407(R)} (\bibinfo {year} {2019})}\BibitemShut {NoStop}%
\bibitem [{\citenamefont {Jugovac}\ \emph {et~al.}(2022)\citenamefont {Jugovac}, \citenamefont {Tresca}, \citenamefont {Cojocariu}, \citenamefont {Di~Santo}, \citenamefont {Zhao}, \citenamefont {Petaccia}, \citenamefont {Moras}, \citenamefont {Profeta},\ and\ \citenamefont {Bisti}}]{Jugovac2022Clarifying}%
  \BibitemOpen
  \bibfield  {author} {\bibinfo {author} {\bibfnamefont {M.}~\bibnamefont {Jugovac}}, \bibinfo {author} {\bibfnamefont {C.}~\bibnamefont {Tresca}}, \bibinfo {author} {\bibfnamefont {I.}~\bibnamefont {Cojocariu}}, \bibinfo {author} {\bibfnamefont {G.}~\bibnamefont {Di~Santo}}, \bibinfo {author} {\bibfnamefont {W.}~\bibnamefont {Zhao}}, \bibinfo {author} {\bibfnamefont {L.}~\bibnamefont {Petaccia}}, \bibinfo {author} {\bibfnamefont {P.}~\bibnamefont {Moras}}, \bibinfo {author} {\bibfnamefont {G.}~\bibnamefont {Profeta}},\ and\ \bibinfo {author} {\bibfnamefont {F.}~\bibnamefont {Bisti}},\ }\href {https://doi.org/10.1103/PhysRevB.105.L241107} {\bibfield  {journal} {\bibinfo  {journal} {Phys. Rev. B}\ }\textbf {\bibinfo {volume} {105}},\ \bibinfo {pages} {L241107} (\bibinfo {year} {2022})}\BibitemShut {NoStop}%
\bibitem [{\citenamefont {Classen}\ and\ \citenamefont {Betouras}(2025)}]{Classen2025High}%
  \BibitemOpen
  \bibfield  {author} {\bibinfo {author} {\bibfnamefont {L.}~\bibnamefont {Classen}}\ and\ \bibinfo {author} {\bibfnamefont {J.~J.}\ \bibnamefont {Betouras}},\ }\href {https://doi.org/https://doi.org/10.1146/annurev-conmatphys-042924-015000} {\bibfield  {journal} {\bibinfo  {journal} {Annual Review of Condensed Matter Physics}\ }\textbf {\bibinfo {volume} {16}},\ \bibinfo {pages} {229} (\bibinfo {year} {2025})}\BibitemShut {NoStop}%
\bibitem [{\citenamefont {Bao}\ \emph {et~al.}(2021)\citenamefont {Bao}, \citenamefont {Zhang}, \citenamefont {Zhang}, \citenamefont {Wu}, \citenamefont {Luo}, \citenamefont {Zhou}, \citenamefont {Li}, \citenamefont {Hou}, \citenamefont {Yao}, \citenamefont {Liu}, \citenamefont {Yu}, \citenamefont {Li}, \citenamefont {Duan}, \citenamefont {Yao}, \citenamefont {Wang},\ and\ \citenamefont {Zhou}}]{Bao2021Experimental}%
  \BibitemOpen
  \bibfield  {author} {\bibinfo {author} {\bibfnamefont {C.}~\bibnamefont {Bao}}, \bibinfo {author} {\bibfnamefont {H.}~\bibnamefont {Zhang}}, \bibinfo {author} {\bibfnamefont {T.}~\bibnamefont {Zhang}}, \bibinfo {author} {\bibfnamefont {X.}~\bibnamefont {Wu}}, \bibinfo {author} {\bibfnamefont {L.}~\bibnamefont {Luo}}, \bibinfo {author} {\bibfnamefont {S.}~\bibnamefont {Zhou}}, \bibinfo {author} {\bibfnamefont {Q.}~\bibnamefont {Li}}, \bibinfo {author} {\bibfnamefont {Y.}~\bibnamefont {Hou}}, \bibinfo {author} {\bibfnamefont {W.}~\bibnamefont {Yao}}, \bibinfo {author} {\bibfnamefont {L.}~\bibnamefont {Liu}}, \bibinfo {author} {\bibfnamefont {P.}~\bibnamefont {Yu}}, \bibinfo {author} {\bibfnamefont {J.}~\bibnamefont {Li}}, \bibinfo {author} {\bibfnamefont {W.}~\bibnamefont {Duan}}, \bibinfo {author} {\bibfnamefont {H.}~\bibnamefont {Yao}}, \bibinfo {author} {\bibfnamefont {Y.}~\bibnamefont {Wang}},\ and\ \bibinfo {author} {\bibfnamefont {S.}~\bibnamefont {Zhou}},\ }\href
  {https://doi.org/10.1103/PhysRevLett.126.206804} {\bibfield  {journal} {\bibinfo  {journal} {Phys. Rev. Lett.}\ }\textbf {\bibinfo {volume} {126}},\ \bibinfo {pages} {206804} (\bibinfo {year} {2021})}\BibitemShut {NoStop}%
\bibitem [{\citenamefont {Dresselhaus}\ and\ \citenamefont {Dresselhaus}(2002)}]{Dresselhaus2002Intercalation}%
  \BibitemOpen
  \bibfield  {author} {\bibinfo {author} {\bibfnamefont {M.~S.}\ \bibnamefont {Dresselhaus}}\ and\ \bibinfo {author} {\bibfnamefont {G.}~\bibnamefont {Dresselhaus}},\ }\href {https://doi.org/10.1080/00018730110113644} {\bibfield  {journal} {\bibinfo  {journal} {Adv. Phys.}\ }\textbf {\bibinfo {volume} {51}},\ \bibinfo {pages} {1} (\bibinfo {year} {2002})}\BibitemShut {NoStop}%
\bibitem [{\citenamefont {Savini}\ \emph {et~al.}(2010)\citenamefont {Savini}, \citenamefont {Ferrari},\ and\ \citenamefont {Giustino}}]{Savini2010first}%
  \BibitemOpen
  \bibfield  {author} {\bibinfo {author} {\bibfnamefont {G.}~\bibnamefont {Savini}}, \bibinfo {author} {\bibfnamefont {A.~C.}\ \bibnamefont {Ferrari}},\ and\ \bibinfo {author} {\bibfnamefont {F.}~\bibnamefont {Giustino}},\ }\href {https://doi.org/10.1103/PhysRevLett.105.037002} {\bibfield  {journal} {\bibinfo  {journal} {Phys. Rev. Lett.}\ }\textbf {\bibinfo {volume} {105}},\ \bibinfo {pages} {037002} (\bibinfo {year} {2010})}\BibitemShut {NoStop}%
\bibitem [{\citenamefont {Profeta}\ \emph {et~al.}(2012)\citenamefont {Profeta}, \citenamefont {Calandra},\ and\ \citenamefont {Mauri}}]{Profeta2012Phonon}%
  \BibitemOpen
  \bibfield  {author} {\bibinfo {author} {\bibfnamefont {G.}~\bibnamefont {Profeta}}, \bibinfo {author} {\bibfnamefont {M.}~\bibnamefont {Calandra}},\ and\ \bibinfo {author} {\bibfnamefont {F.}~\bibnamefont {Mauri}},\ }\href {https://doi.org/10.1038/nphys2181} {\bibfield  {journal} {\bibinfo  {journal} {Nat. Phys.}\ }\textbf {\bibinfo {volume} {8}},\ \bibinfo {pages} {131} (\bibinfo {year} {2012})}\BibitemShut {NoStop}%
\bibitem [{\citenamefont {Chapman}\ \emph {et~al.}(2016)\citenamefont {Chapman}, \citenamefont {Su}, \citenamefont {Howard}, \citenamefont {Kundys}, \citenamefont {Grigorenko}, \citenamefont {Guinea}, \citenamefont {Geim}, \citenamefont {Grigorieva},\ and\ \citenamefont {Nair}}]{Chapman2016Superconductivity}%
  \BibitemOpen
  \bibfield  {author} {\bibinfo {author} {\bibfnamefont {J.}~\bibnamefont {Chapman}}, \bibinfo {author} {\bibfnamefont {Y.}~\bibnamefont {Su}}, \bibinfo {author} {\bibfnamefont {C.~A.}\ \bibnamefont {Howard}}, \bibinfo {author} {\bibfnamefont {D.}~\bibnamefont {Kundys}}, \bibinfo {author} {\bibfnamefont {A.~N.}\ \bibnamefont {Grigorenko}}, \bibinfo {author} {\bibfnamefont {F.}~\bibnamefont {Guinea}}, \bibinfo {author} {\bibfnamefont {A.~K.}\ \bibnamefont {Geim}}, \bibinfo {author} {\bibfnamefont {I.~V.}\ \bibnamefont {Grigorieva}},\ and\ \bibinfo {author} {\bibfnamefont {R.~R.}\ \bibnamefont {Nair}},\ }\href {https://doi.org/10.1038/srep23254} {\bibfield  {journal} {\bibinfo  {journal} {Sci. Rep.}\ }\textbf {\bibinfo {volume} {6}},\ \bibinfo {pages} {23254} (\bibinfo {year} {2016})}\BibitemShut {NoStop}%
\bibitem [{\citenamefont {Ludbrook}\ \emph {et~al.}(2015)\citenamefont {Ludbrook}, \citenamefont {Levy}, \citenamefont {Nigge}, \citenamefont {Zonno}, \citenamefont {Schneider}, \citenamefont {Dvorak}, \citenamefont {Veenstra}, \citenamefont {Zhdanovich}, \citenamefont {Wong}, \citenamefont {Dosanjh}, \citenamefont {Straßer}, \citenamefont {St\"ohr}, \citenamefont {Forti}, \citenamefont {Ast}, \citenamefont {Starke},\ and\ \citenamefont {Damascelli}}]{Ludbrook2015Evidence}%
  \BibitemOpen
  \bibfield  {author} {\bibinfo {author} {\bibfnamefont {B.~M.}\ \bibnamefont {Ludbrook}}, \bibinfo {author} {\bibfnamefont {G.}~\bibnamefont {Levy}}, \bibinfo {author} {\bibfnamefont {P.}~\bibnamefont {Nigge}}, \bibinfo {author} {\bibfnamefont {M.}~\bibnamefont {Zonno}}, \bibinfo {author} {\bibfnamefont {M.}~\bibnamefont {Schneider}}, \bibinfo {author} {\bibfnamefont {D.~J.}\ \bibnamefont {Dvorak}}, \bibinfo {author} {\bibfnamefont {C.~N.}\ \bibnamefont {Veenstra}}, \bibinfo {author} {\bibfnamefont {S.}~\bibnamefont {Zhdanovich}}, \bibinfo {author} {\bibfnamefont {D.}~\bibnamefont {Wong}}, \bibinfo {author} {\bibfnamefont {P.}~\bibnamefont {Dosanjh}}, \bibinfo {author} {\bibfnamefont {C.}~\bibnamefont {Straßer}}, \bibinfo {author} {\bibfnamefont {A.}~\bibnamefont {St\"ohr}}, \bibinfo {author} {\bibfnamefont {S.}~\bibnamefont {Forti}}, \bibinfo {author} {\bibfnamefont {C.~R.}\ \bibnamefont {Ast}}, \bibinfo {author} {\bibfnamefont {U.}~\bibnamefont {Starke}},\ and\ \bibinfo {author} {\bibfnamefont
  {A.}~\bibnamefont {Damascelli}},\ }\href {https://doi.org/10.1073/pnas.1510435112} {\bibfield  {journal} {\bibinfo  {journal} {Proc. Natl. Acad. Sci. U.S.A.}\ }\textbf {\bibinfo {volume} {112}},\ \bibinfo {pages} {11795} (\bibinfo {year} {2015})}\BibitemShut {NoStop}%
\bibitem [{\citenamefont {Wu}\ and\ \citenamefont {Hu}(2016)}]{Wu2016Topological}%
  \BibitemOpen
  \bibfield  {author} {\bibinfo {author} {\bibfnamefont {L.-H.}\ \bibnamefont {Wu}}\ and\ \bibinfo {author} {\bibfnamefont {X.}~\bibnamefont {Hu}},\ }\href {https://doi.org/10.1038/srep24347} {\bibfield  {journal} {\bibinfo  {journal} {Scientific Reports}\ }\textbf {\bibinfo {volume} {6}},\ \bibinfo {pages} {24347} (\bibinfo {year} {2016})}\BibitemShut {NoStop}%
\bibitem [{\citenamefont {Freeney}\ \emph {et~al.}(2020)\citenamefont {Freeney}, \citenamefont {van~den Broeke}, \citenamefont {Harsveld van~der Veen}, \citenamefont {Swart},\ and\ \citenamefont {Morais~Smith}}]{Freeney2020Edge}%
  \BibitemOpen
  \bibfield  {author} {\bibinfo {author} {\bibfnamefont {S.~E.}\ \bibnamefont {Freeney}}, \bibinfo {author} {\bibfnamefont {J.~J.}\ \bibnamefont {van~den Broeke}}, \bibinfo {author} {\bibfnamefont {A.~J.~J.}\ \bibnamefont {Harsveld van~der Veen}}, \bibinfo {author} {\bibfnamefont {I.}~\bibnamefont {Swart}},\ and\ \bibinfo {author} {\bibfnamefont {C.}~\bibnamefont {Morais~Smith}},\ }\href {https://doi.org/10.1103/PhysRevLett.124.236404} {\bibfield  {journal} {\bibinfo  {journal} {Phys. Rev. Lett.}\ }\textbf {\bibinfo {volume} {124}},\ \bibinfo {pages} {236404} (\bibinfo {year} {2020})}\BibitemShut {NoStop}%
\bibitem [{\citenamefont {García}\ \emph {et~al.}(2024)\citenamefont {García}, \citenamefont {Betancur-Ocampo}, \citenamefont {Sánchez-Ochoa},\ and\ \citenamefont {Stegmann}}]{Garcia2024Atomically}%
  \BibitemOpen
  \bibfield  {author} {\bibinfo {author} {\bibfnamefont {S.~G.~y.}\ \bibnamefont {García}}, \bibinfo {author} {\bibfnamefont {Y.}~\bibnamefont {Betancur-Ocampo}}, \bibinfo {author} {\bibfnamefont {F.}~\bibnamefont {Sánchez-Ochoa}},\ and\ \bibinfo {author} {\bibfnamefont {T.}~\bibnamefont {Stegmann}},\ }\href {https://doi.org/10.1021/acs.nanolett.3c04703} {\bibfield  {journal} {\bibinfo  {journal} {Nano Letters}\ }\textbf {\bibinfo {volume} {24}},\ \bibinfo {pages} {2322} (\bibinfo {year} {2024})},\ \bibinfo {note} {pMID: 38329068},\ \Eprint {https://arxiv.org/abs/https://doi.org/10.1021/acs.nanolett.3c04703} {https://doi.org/10.1021/acs.nanolett.3c04703} \BibitemShut {NoStop}%
\bibitem [{\citenamefont {Herrera}\ \emph {et~al.}(2024)\citenamefont {Herrera}, \citenamefont {Parra-Martínez}, \citenamefont {Rosenzweig}, \citenamefont {Matta}, \citenamefont {Polley}, \citenamefont {K{\"u}ster}, \citenamefont {Starke}, \citenamefont {Guinea}, \citenamefont {Silva-Guill{\'e}n}, \citenamefont {Naumis},\ and\ \citenamefont {Pantaleón}}]{Herrera2024Topological}%
  \BibitemOpen
  \bibfield  {author} {\bibinfo {author} {\bibfnamefont {S.~A.}\ \bibnamefont {Herrera}}, \bibinfo {author} {\bibfnamefont {G.}~\bibnamefont {Parra-Martínez}}, \bibinfo {author} {\bibfnamefont {P.}~\bibnamefont {Rosenzweig}}, \bibinfo {author} {\bibfnamefont {B.}~\bibnamefont {Matta}}, \bibinfo {author} {\bibfnamefont {C.~M.}\ \bibnamefont {Polley}}, \bibinfo {author} {\bibfnamefont {K.}~\bibnamefont {K{\"u}ster}}, \bibinfo {author} {\bibfnamefont {U.}~\bibnamefont {Starke}}, \bibinfo {author} {\bibfnamefont {F.}~\bibnamefont {Guinea}}, \bibinfo {author} {\bibfnamefont {J.~Ã.}\ \bibnamefont {Silva-Guill{\'e}n}}, \bibinfo {author} {\bibfnamefont {G.~G.}\ \bibnamefont {Naumis}},\ and\ \bibinfo {author} {\bibfnamefont {P.~A.}\ \bibnamefont {Pantaleón}},\ }\href {https://doi.org/10.1021/acsnano.4c12532} {\bibfield  {journal} {\bibinfo  {journal} {ACS Nano}\ }\textbf {\bibinfo {volume} {18}},\ \bibinfo {pages} {34842} (\bibinfo {year} {2024})},\ \bibinfo {note} {pMID: 39652458},\ \Eprint
  {https://arxiv.org/abs/https://doi.org/10.1021/acsnano.4c12532} {https://doi.org/10.1021/acsnano.4c12532} \BibitemShut {NoStop}%
\bibitem [{\citenamefont {Petrovi{\'c}}\ \emph {et~al.}(2013)\citenamefont {Petrovi{\'c}}, \citenamefont {{\v S}rut~Raki{\'c}}, \citenamefont {Runte}, \citenamefont {Busse}, \citenamefont {Sadowski}, \citenamefont {Lazi{\'c}}, \citenamefont {Pletikosi{\'c}}, \citenamefont {Pan}, \citenamefont {Milun}, \citenamefont {Pervan}, \citenamefont {Atodiresei}, \citenamefont {Brako}, \citenamefont {{\v S}ok{\v c}evi{\'c}}, \citenamefont {Valla}, \citenamefont {Michely},\ and\ \citenamefont {Kralj}}]{Petrovic2013the}%
  \BibitemOpen
  \bibfield  {author} {\bibinfo {author} {\bibfnamefont {M.}~\bibnamefont {Petrovi{\'c}}}, \bibinfo {author} {\bibfnamefont {I.}~\bibnamefont {{\v S}rut~Raki{\'c}}}, \bibinfo {author} {\bibfnamefont {S.}~\bibnamefont {Runte}}, \bibinfo {author} {\bibfnamefont {C.}~\bibnamefont {Busse}}, \bibinfo {author} {\bibfnamefont {J.~T.}\ \bibnamefont {Sadowski}}, \bibinfo {author} {\bibfnamefont {P.}~\bibnamefont {Lazi{\'c}}}, \bibinfo {author} {\bibfnamefont {I.}~\bibnamefont {Pletikosi{\'c}}}, \bibinfo {author} {\bibfnamefont {Z.~H.}\ \bibnamefont {Pan}}, \bibinfo {author} {\bibfnamefont {M.}~\bibnamefont {Milun}}, \bibinfo {author} {\bibfnamefont {P.}~\bibnamefont {Pervan}}, \bibinfo {author} {\bibfnamefont {N.}~\bibnamefont {Atodiresei}}, \bibinfo {author} {\bibfnamefont {R.}~\bibnamefont {Brako}}, \bibinfo {author} {\bibfnamefont {D.}~\bibnamefont {{\v S}ok{\v c}evi{\'c}}}, \bibinfo {author} {\bibfnamefont {T.}~\bibnamefont {Valla}}, \bibinfo {author} {\bibfnamefont {T.}~\bibnamefont {Michely}},\ and\ \bibinfo
  {author} {\bibfnamefont {M.}~\bibnamefont {Kralj}},\ }\href {https://doi.org/10.1038/ncomms3772} {\bibfield  {journal} {\bibinfo  {journal} {Nature Communications}\ }\textbf {\bibinfo {volume} {4}},\ \bibinfo {pages} {2772} (\bibinfo {year} {2013})}\BibitemShut {NoStop}%
\bibitem [{\citenamefont {Hell}\ \emph {et~al.}(2020)\citenamefont {Hell}, \citenamefont {Ehlen}, \citenamefont {Marini}, \citenamefont {Falke}, \citenamefont {Senkovskiy}, \citenamefont {Herbig}, \citenamefont {Teichert}, \citenamefont {Jolie}, \citenamefont {Michely}, \citenamefont {Avila}, \citenamefont {Santo}, \citenamefont {Torre}, \citenamefont {Petaccia}, \citenamefont {Profeta},\ and\ \citenamefont {Gr{\"u}neis}}]{Hell2020massive}%
  \BibitemOpen
  \bibfield  {author} {\bibinfo {author} {\bibfnamefont {M.}~\bibnamefont {Hell}}, \bibinfo {author} {\bibfnamefont {N.}~\bibnamefont {Ehlen}}, \bibinfo {author} {\bibfnamefont {G.}~\bibnamefont {Marini}}, \bibinfo {author} {\bibfnamefont {Y.}~\bibnamefont {Falke}}, \bibinfo {author} {\bibfnamefont {B.~V.}\ \bibnamefont {Senkovskiy}}, \bibinfo {author} {\bibfnamefont {C.}~\bibnamefont {Herbig}}, \bibinfo {author} {\bibfnamefont {C.}~\bibnamefont {Teichert}}, \bibinfo {author} {\bibfnamefont {W.}~\bibnamefont {Jolie}}, \bibinfo {author} {\bibfnamefont {T.}~\bibnamefont {Michely}}, \bibinfo {author} {\bibfnamefont {J.}~\bibnamefont {Avila}}, \bibinfo {author} {\bibfnamefont {G.~D.}\ \bibnamefont {Santo}}, \bibinfo {author} {\bibfnamefont {D.~M. d.~l.}\ \bibnamefont {Torre}}, \bibinfo {author} {\bibfnamefont {L.}~\bibnamefont {Petaccia}}, \bibinfo {author} {\bibfnamefont {G.}~\bibnamefont {Profeta}},\ and\ \bibinfo {author} {\bibfnamefont {A.}~\bibnamefont {Gr{\"u}neis}},\ }\href
  {https://doi.org/10.1038/s41467-020-15130-1} {\bibfield  {journal} {\bibinfo  {journal} {Nature Communications}\ }\textbf {\bibinfo {volume} {11}},\ \bibinfo {pages} {1340} (\bibinfo {year} {2020})}\BibitemShut {NoStop}%
\bibitem [{\citenamefont {Sugawara}\ \emph {et~al.}(2011)\citenamefont {Sugawara}, \citenamefont {Kanetani}, \citenamefont {Sato},\ and\ \citenamefont {Takahashi}}]{Sugawara2011Fabrication}%
  \BibitemOpen
  \bibfield  {author} {\bibinfo {author} {\bibfnamefont {K.}~\bibnamefont {Sugawara}}, \bibinfo {author} {\bibfnamefont {K.}~\bibnamefont {Kanetani}}, \bibinfo {author} {\bibfnamefont {T.}~\bibnamefont {Sato}},\ and\ \bibinfo {author} {\bibfnamefont {T.}~\bibnamefont {Takahashi}},\ }\href {https://doi.org/10.1063/1.3582814} {\bibfield  {journal} {\bibinfo  {journal} {AIP Adv.}\ }\textbf {\bibinfo {volume} {1}},\ \bibinfo {pages} {022103} (\bibinfo {year} {2011})}\BibitemShut {NoStop}%
\bibitem [{\citenamefont {Ichinokura}\ \emph {et~al.}(2022)\citenamefont {Ichinokura}, \citenamefont {Toyoda}, \citenamefont {Hashizume}, \citenamefont {Horii}, \citenamefont {Kusaka}, \citenamefont {Ideta}, \citenamefont {Tanaka}, \citenamefont {Shimizu}, \citenamefont {Hitosugi}, \citenamefont {Saito},\ and\ \citenamefont {Hirahara}}]{Ichinokura2022Van}%
  \BibitemOpen
  \bibfield  {author} {\bibinfo {author} {\bibfnamefont {S.}~\bibnamefont {Ichinokura}}, \bibinfo {author} {\bibfnamefont {M.}~\bibnamefont {Toyoda}}, \bibinfo {author} {\bibfnamefont {M.}~\bibnamefont {Hashizume}}, \bibinfo {author} {\bibfnamefont {K.}~\bibnamefont {Horii}}, \bibinfo {author} {\bibfnamefont {S.}~\bibnamefont {Kusaka}}, \bibinfo {author} {\bibfnamefont {S.}~\bibnamefont {Ideta}}, \bibinfo {author} {\bibfnamefont {K.}~\bibnamefont {Tanaka}}, \bibinfo {author} {\bibfnamefont {R.}~\bibnamefont {Shimizu}}, \bibinfo {author} {\bibfnamefont {T.}~\bibnamefont {Hitosugi}}, \bibinfo {author} {\bibfnamefont {S.}~\bibnamefont {Saito}},\ and\ \bibinfo {author} {\bibfnamefont {T.}~\bibnamefont {Hirahara}},\ }\href {https://doi.org/10.1103/PhysRevB.105.235307} {\bibfield  {journal} {\bibinfo  {journal} {Phys. Rev. B}\ }\textbf {\bibinfo {volume} {105}},\ \bibinfo {pages} {235307} (\bibinfo {year} {2022})}\BibitemShut {NoStop}%
\bibitem [{\citenamefont {Wu}\ \emph {et~al.}(2023)\citenamefont {Wu}, \citenamefont {Zheng}, \citenamefont {Kang},\ and\ \citenamefont {Li}}]{Wu2023Effects}%
  \BibitemOpen
  \bibfield  {author} {\bibinfo {author} {\bibfnamefont {X.}~\bibnamefont {Wu}}, \bibinfo {author} {\bibfnamefont {F.}~\bibnamefont {Zheng}}, \bibinfo {author} {\bibfnamefont {F.}~\bibnamefont {Kang}},\ and\ \bibinfo {author} {\bibfnamefont {J.}~\bibnamefont {Li}},\ }\href {https://doi.org/10.1103/PhysRevB.107.165409} {\bibfield  {journal} {\bibinfo  {journal} {Phys. Rev. B}\ }\textbf {\bibinfo {volume} {107}},\ \bibinfo {pages} {165409} (\bibinfo {year} {2023})}\BibitemShut {NoStop}%
\bibitem [{\citenamefont {Cs{\'a}nyi}\ \emph {et~al.}(2005)\citenamefont {Cs{\'a}nyi}, \citenamefont {Littlewood}, \citenamefont {Nevidomskyy}, \citenamefont {Pickard},\ and\ \citenamefont {Simons}}]{Csanyi2005}%
  \BibitemOpen
  \bibfield  {author} {\bibinfo {author} {\bibfnamefont {G.}~\bibnamefont {Cs{\'a}nyi}}, \bibinfo {author} {\bibfnamefont {P.~B.}\ \bibnamefont {Littlewood}}, \bibinfo {author} {\bibfnamefont {A.~H.}\ \bibnamefont {Nevidomskyy}}, \bibinfo {author} {\bibfnamefont {C.~J.}\ \bibnamefont {Pickard}},\ and\ \bibinfo {author} {\bibfnamefont {B.~D.}\ \bibnamefont {Simons}},\ }\href {https://doi.org/10.1038/nphys119} {\bibfield  {journal} {\bibinfo  {journal} {Nat. Phys.}\ }\textbf {\bibinfo {volume} {1}},\ \bibinfo {pages} {42} (\bibinfo {year} {2005})}\BibitemShut {NoStop}%
\bibitem [{\citenamefont {Kanetani}\ \emph {et~al.}(2012)\citenamefont {Kanetani}, \citenamefont {Sugawara}, \citenamefont {Sato}, \citenamefont {Shimizu}, \citenamefont {Iwaya}, \citenamefont {Hitosugi},\ and\ \citenamefont {Takahashi}}]{Kanetani2012ca}%
  \BibitemOpen
  \bibfield  {author} {\bibinfo {author} {\bibfnamefont {K.}~\bibnamefont {Kanetani}}, \bibinfo {author} {\bibfnamefont {K.}~\bibnamefont {Sugawara}}, \bibinfo {author} {\bibfnamefont {T.}~\bibnamefont {Sato}}, \bibinfo {author} {\bibfnamefont {R.}~\bibnamefont {Shimizu}}, \bibinfo {author} {\bibfnamefont {K.}~\bibnamefont {Iwaya}}, \bibinfo {author} {\bibfnamefont {T.}~\bibnamefont {Hitosugi}},\ and\ \bibinfo {author} {\bibfnamefont {T.}~\bibnamefont {Takahashi}},\ }\href {https://doi.org/10.1073/pnas.1208889109} {\bibfield  {journal} {\bibinfo  {journal} {Proceedings of the National Academy of Sciences}\ }\textbf {\bibinfo {volume} {109}},\ \bibinfo {pages} {19610} (\bibinfo {year} {2012})},\ \Eprint {https://arxiv.org/abs/https://www.pnas.org/doi/pdf/10.1073/pnas.1208889109} {https://www.pnas.org/doi/pdf/10.1073/pnas.1208889109} \BibitemShut {NoStop}%
\bibitem [{\citenamefont {Huempfner}\ \emph {et~al.}(2023)\citenamefont {Huempfner}, \citenamefont {Otto}, \citenamefont {Forker}, \citenamefont {Müller},\ and\ \citenamefont {Fritz}}]{Huempfner2023Superconductivity}%
  \BibitemOpen
  \bibfield  {author} {\bibinfo {author} {\bibfnamefont {T.}~\bibnamefont {Huempfner}}, \bibinfo {author} {\bibfnamefont {F.}~\bibnamefont {Otto}}, \bibinfo {author} {\bibfnamefont {R.}~\bibnamefont {Forker}}, \bibinfo {author} {\bibfnamefont {P.}~\bibnamefont {Müller}},\ and\ \bibinfo {author} {\bibfnamefont {T.}~\bibnamefont {Fritz}},\ }\href {https://doi.org/10.1002/admi.202300014} {\bibfield  {journal} {\bibinfo  {journal} {Adv. Mater. Interfaces}\ }\textbf {\bibinfo {volume} {10}},\ \bibinfo {pages} {2300014} (\bibinfo {year} {2023})}\BibitemShut {NoStop}%
\bibitem [{\citenamefont {Kohn}\ and\ \citenamefont {Sham}(1965)}]{kohsha1965}%
  \BibitemOpen
  \bibfield  {author} {\bibinfo {author} {\bibfnamefont {W.}~\bibnamefont {Kohn}}\ and\ \bibinfo {author} {\bibfnamefont {L.~J.}\ \bibnamefont {Sham}},\ }\href {https://doi.org/10.1103/PhysRev.140.A1133} {\bibfield  {journal} {\bibinfo  {journal} {Phys. Rev.}\ }\textbf {\bibinfo {volume} {140}},\ \bibinfo {pages} {A1133} (\bibinfo {year} {1965})}\BibitemShut {NoStop}%
\bibitem [{\citenamefont {Hohenberg}\ and\ \citenamefont {Kohn}(1964)}]{HohKoh1964}%
  \BibitemOpen
  \bibfield  {author} {\bibinfo {author} {\bibfnamefont {P.}~\bibnamefont {Hohenberg}}\ and\ \bibinfo {author} {\bibfnamefont {W.}~\bibnamefont {Kohn}},\ }\href {https://doi.org/10.1103/PhysRev.136.B864} {\bibfield  {journal} {\bibinfo  {journal} {Phys. Rev.}\ }\textbf {\bibinfo {volume} {136}},\ \bibinfo {pages} {B864} (\bibinfo {year} {1964})}\BibitemShut {NoStop}%
\bibitem [{\citenamefont {Artacho}\ \emph {et~al.}(2008)\citenamefont {Artacho}, \citenamefont {Anglada}, \citenamefont {Di\'eguez}, \citenamefont {Gale}, \citenamefont {Garc\'ia}, \citenamefont {Junquera}, \citenamefont {Martin}, \citenamefont {Ordej\'on}, \citenamefont {Pruneda}, \citenamefont {S\'anchez-Portal},\ and\ \citenamefont {Soler}}]{ArtAng2008}%
  \BibitemOpen
  \bibfield  {author} {\bibinfo {author} {\bibfnamefont {E.}~\bibnamefont {Artacho}}, \bibinfo {author} {\bibfnamefont {E.}~\bibnamefont {Anglada}}, \bibinfo {author} {\bibfnamefont {O.}~\bibnamefont {Di\'eguez}}, \bibinfo {author} {\bibfnamefont {J.~D.}\ \bibnamefont {Gale}}, \bibinfo {author} {\bibfnamefont {A.}~\bibnamefont {Garc\'ia}}, \bibinfo {author} {\bibfnamefont {J.}~\bibnamefont {Junquera}}, \bibinfo {author} {\bibfnamefont {R.~M.}\ \bibnamefont {Martin}}, \bibinfo {author} {\bibfnamefont {P.}~\bibnamefont {Ordej\'on}}, \bibinfo {author} {\bibfnamefont {J.~M.}\ \bibnamefont {Pruneda}}, \bibinfo {author} {\bibfnamefont {D.}~\bibnamefont {S\'anchez-Portal}},\ and\ \bibinfo {author} {\bibfnamefont {J.~M.}\ \bibnamefont {Soler}},\ }\href {https://doi.org/10.1088/0953-8984/20/6/064208} {\bibfield  {journal} {\bibinfo  {journal} {J. Phys.: Condens. Matter}\ }\textbf {\bibinfo {volume} {20}},\ \bibinfo {pages} {064208} (\bibinfo {year} {2008})}\BibitemShut {NoStop}%
\bibitem [{\citenamefont {Soler}\ \emph {et~al.}(2002)\citenamefont {Soler}, \citenamefont {Artacho}, \citenamefont {Gale}, \citenamefont {Garc\'ia}, \citenamefont {Junquera}, \citenamefont {Ordej\'on},\ and\ \citenamefont {S\'anchez-Portal}}]{SolArt2002}%
  \BibitemOpen
  \bibfield  {author} {\bibinfo {author} {\bibfnamefont {J.~M.}\ \bibnamefont {Soler}}, \bibinfo {author} {\bibfnamefont {E.}~\bibnamefont {Artacho}}, \bibinfo {author} {\bibfnamefont {J.~D.}\ \bibnamefont {Gale}}, \bibinfo {author} {\bibfnamefont {A.}~\bibnamefont {Garc\'ia}}, \bibinfo {author} {\bibfnamefont {J.}~\bibnamefont {Junquera}}, \bibinfo {author} {\bibfnamefont {P.}~\bibnamefont {Ordej\'on}},\ and\ \bibinfo {author} {\bibfnamefont {D.}~\bibnamefont {S\'anchez-Portal}},\ }\href {https://doi.org/10.1088/0953-8984/14/11/302} {\bibfield  {journal} {\bibinfo  {journal} {J. Phys.: Condens. Matter}\ }\textbf {\bibinfo {volume} {14}},\ \bibinfo {pages} {2745} (\bibinfo {year} {2002})}\BibitemShut {NoStop}%
\bibitem [{\citenamefont {García}\ \emph {et~al.}(2020)\citenamefont {García}, \citenamefont {Papior}, \citenamefont {Akhtar}, \citenamefont {Artacho}, \citenamefont {Blum}, \citenamefont {Bosoni}, \citenamefont {Brandimarte}, \citenamefont {Brandbyge}, \citenamefont {Cerdá}, \citenamefont {Corsetti}, \citenamefont {Cuadrado}, \citenamefont {Dikan}, \citenamefont {Ferrer}, \citenamefont {Gale}, \citenamefont {García-Fernández}, \citenamefont {García-Suárez}, \citenamefont {García}, \citenamefont {Huhs}, \citenamefont {Illera}, \citenamefont {Korytár}, \citenamefont {Koval}, \citenamefont {Lebedeva}, \citenamefont {Lin}, \citenamefont {López-Tarifa}, \citenamefont {Mayo}, \citenamefont {Mohr}, \citenamefont {Ordejón}, \citenamefont {Postnikov}, \citenamefont {Pouillon}, \citenamefont {Pruneda}, \citenamefont {Robles}, \citenamefont {Sánchez-Portal}, \citenamefont {Soler}, \citenamefont {Ullah}, \citenamefont {Yu},\ and\ \citenamefont {Junquera}}]{siesta-2020}%
  \BibitemOpen
  \bibfield  {author} {\bibinfo {author} {\bibfnamefont {A.}~\bibnamefont {García}}, \bibinfo {author} {\bibfnamefont {N.}~\bibnamefont {Papior}}, \bibinfo {author} {\bibfnamefont {A.}~\bibnamefont {Akhtar}}, \bibinfo {author} {\bibfnamefont {E.}~\bibnamefont {Artacho}}, \bibinfo {author} {\bibfnamefont {V.}~\bibnamefont {Blum}}, \bibinfo {author} {\bibfnamefont {E.}~\bibnamefont {Bosoni}}, \bibinfo {author} {\bibfnamefont {P.}~\bibnamefont {Brandimarte}}, \bibinfo {author} {\bibfnamefont {M.}~\bibnamefont {Brandbyge}}, \bibinfo {author} {\bibfnamefont {J.~I.}\ \bibnamefont {Cerdá}}, \bibinfo {author} {\bibfnamefont {F.}~\bibnamefont {Corsetti}}, \bibinfo {author} {\bibfnamefont {R.}~\bibnamefont {Cuadrado}}, \bibinfo {author} {\bibfnamefont {V.}~\bibnamefont {Dikan}}, \bibinfo {author} {\bibfnamefont {J.}~\bibnamefont {Ferrer}}, \bibinfo {author} {\bibfnamefont {J.}~\bibnamefont {Gale}}, \bibinfo {author} {\bibfnamefont {P.}~\bibnamefont {García-Fernández}}, \bibinfo {author} {\bibfnamefont {V.~M.}\
  \bibnamefont {García-Suárez}}, \bibinfo {author} {\bibfnamefont {S.}~\bibnamefont {García}}, \bibinfo {author} {\bibfnamefont {G.}~\bibnamefont {Huhs}}, \bibinfo {author} {\bibfnamefont {S.}~\bibnamefont {Illera}}, \bibinfo {author} {\bibfnamefont {R.}~\bibnamefont {Korytár}}, \bibinfo {author} {\bibfnamefont {P.}~\bibnamefont {Koval}}, \bibinfo {author} {\bibfnamefont {I.}~\bibnamefont {Lebedeva}}, \bibinfo {author} {\bibfnamefont {L.}~\bibnamefont {Lin}}, \bibinfo {author} {\bibfnamefont {P.}~\bibnamefont {López-Tarifa}}, \bibinfo {author} {\bibfnamefont {S.~G.}\ \bibnamefont {Mayo}}, \bibinfo {author} {\bibfnamefont {S.}~\bibnamefont {Mohr}}, \bibinfo {author} {\bibfnamefont {P.}~\bibnamefont {Ordejón}}, \bibinfo {author} {\bibfnamefont {A.}~\bibnamefont {Postnikov}}, \bibinfo {author} {\bibfnamefont {Y.}~\bibnamefont {Pouillon}}, \bibinfo {author} {\bibfnamefont {M.}~\bibnamefont {Pruneda}}, \bibinfo {author} {\bibfnamefont {R.}~\bibnamefont {Robles}}, \bibinfo {author} {\bibfnamefont
  {D.}~\bibnamefont {Sánchez-Portal}}, \bibinfo {author} {\bibfnamefont {J.~M.}\ \bibnamefont {Soler}}, \bibinfo {author} {\bibfnamefont {R.}~\bibnamefont {Ullah}}, \bibinfo {author} {\bibfnamefont {V.~W.-z.}\ \bibnamefont {Yu}},\ and\ \bibinfo {author} {\bibfnamefont {J.}~\bibnamefont {Junquera}},\ }\href {https://doi.org/10.1063/5.0005077} {\bibfield  {journal} {\bibinfo  {journal} {J. Chem. Phys.}\ }\textbf {\bibinfo {volume} {152}},\ \bibinfo {pages} {204108} (\bibinfo {year} {2020})}\BibitemShut {NoStop}%
\bibitem [{\citenamefont {{van Setten}}\ \emph {et~al.}(2018)\citenamefont {{van Setten}}, \citenamefont {Giantomassi}, \citenamefont {Bousquet}, \citenamefont {Verstraete}, \citenamefont {Hamann}, \citenamefont {Gonze},\ and\ \citenamefont {Rignanese}}]{vansetten2018dojo}%
  \BibitemOpen
  \bibfield  {author} {\bibinfo {author} {\bibfnamefont {M.}~\bibnamefont {{van Setten}}}, \bibinfo {author} {\bibfnamefont {M.}~\bibnamefont {Giantomassi}}, \bibinfo {author} {\bibfnamefont {E.}~\bibnamefont {Bousquet}}, \bibinfo {author} {\bibfnamefont {M.}~\bibnamefont {Verstraete}}, \bibinfo {author} {\bibfnamefont {D.}~\bibnamefont {Hamann}}, \bibinfo {author} {\bibfnamefont {X.}~\bibnamefont {Gonze}},\ and\ \bibinfo {author} {\bibfnamefont {G.-M.}\ \bibnamefont {Rignanese}},\ }\href {https://doi.org/10.1016/j.cpc.2018.01.012} {\bibfield  {journal} {\bibinfo  {journal} {Comput. Phys. Commun.}\ }\textbf {\bibinfo {volume} {226}},\ \bibinfo {pages} {39} (\bibinfo {year} {2018})}\BibitemShut {NoStop}%
\bibitem [{\citenamefont {García}\ \emph {et~al.}(2018)\citenamefont {García}, \citenamefont {Verstraete}, \citenamefont {Pouillon},\ and\ \citenamefont {Junquera}}]{garcia2018psml}%
  \BibitemOpen
  \bibfield  {author} {\bibinfo {author} {\bibfnamefont {A.}~\bibnamefont {García}}, \bibinfo {author} {\bibfnamefont {M.~J.}\ \bibnamefont {Verstraete}}, \bibinfo {author} {\bibfnamefont {Y.}~\bibnamefont {Pouillon}},\ and\ \bibinfo {author} {\bibfnamefont {J.}~\bibnamefont {Junquera}},\ }\href {https://doi.org/10.1016/j.cpc.2018.02.011} {\bibfield  {journal} {\bibinfo  {journal} {Comput. Phys. Commun.}\ }\textbf {\bibinfo {volume} {227}},\ \bibinfo {pages} {51} (\bibinfo {year} {2018})}\BibitemShut {NoStop}%
\bibitem [{\citenamefont {Perdew}\ \emph {et~al.}(1996)\citenamefont {Perdew}, \citenamefont {Burke},\ and\ \citenamefont {Ernzerhof}}]{PBE96}%
  \BibitemOpen
  \bibfield  {author} {\bibinfo {author} {\bibfnamefont {J.~P.}\ \bibnamefont {Perdew}}, \bibinfo {author} {\bibfnamefont {K.}~\bibnamefont {Burke}},\ and\ \bibinfo {author} {\bibfnamefont {M.}~\bibnamefont {Ernzerhof}},\ }\href {https://doi.org/10.1103/PhysRevLett.77.3865} {\bibfield  {journal} {\bibinfo  {journal} {Phys. Rev. Lett.}\ }\textbf {\bibinfo {volume} {77}},\ \bibinfo {pages} {3865} (\bibinfo {year} {1996})}\BibitemShut {NoStop}%
\bibitem [{\citenamefont {Grimme}(2006)}]{grimme2006semiempirical}%
  \BibitemOpen
  \bibfield  {author} {\bibinfo {author} {\bibfnamefont {S.}~\bibnamefont {Grimme}},\ }\href {https://doi.org/10.1002/jcc.20495} {\bibfield  {journal} {\bibinfo  {journal} {J. Comput. Chem.}\ }\textbf {\bibinfo {volume} {27}},\ \bibinfo {pages} {1787} (\bibinfo {year} {2006})}\BibitemShut {NoStop}%
\bibitem [{\citenamefont {Artacho}\ \emph {et~al.}(1999)\citenamefont {Artacho}, \citenamefont {S\'{a}nchez-Portal}, \citenamefont {Ordej\'{o}n}, \citenamefont {Garc\'{i}a},\ and\ \citenamefont {Soler}}]{arsan99}%
  \BibitemOpen
  \bibfield  {author} {\bibinfo {author} {\bibfnamefont {E.}~\bibnamefont {Artacho}}, \bibinfo {author} {\bibfnamefont {D.}~\bibnamefont {S\'{a}nchez-Portal}}, \bibinfo {author} {\bibfnamefont {P.}~\bibnamefont {Ordej\'{o}n}}, \bibinfo {author} {\bibfnamefont {A.}~\bibnamefont {Garc\'{i}a}},\ and\ \bibinfo {author} {\bibfnamefont {J.~M.}\ \bibnamefont {Soler}},\ }\href {https://doi.org/10.1002/(SICI)1521-3951(199909)215:1<809::AID-PSSB809>3.0.CO;2-0} {\bibfield  {journal} {\bibinfo  {journal} {Phys. Status Solidi (B)}\ }\textbf {\bibinfo {volume} {215}},\ \bibinfo {pages} {809} (\bibinfo {year} {1999})}\BibitemShut {NoStop}%
\bibitem [{\citenamefont {Monkhorst}\ and\ \citenamefont {Pack}(1976)}]{MonPac76}%
  \BibitemOpen
  \bibfield  {author} {\bibinfo {author} {\bibfnamefont {H.~J.}\ \bibnamefont {Monkhorst}}\ and\ \bibinfo {author} {\bibfnamefont {J.~D.}\ \bibnamefont {Pack}},\ }\href {https://doi.org/10.1103/PhysRevB.13.5188} {\bibfield  {journal} {\bibinfo  {journal} {Phys. Rev. B}\ }\textbf {\bibinfo {volume} {13}},\ \bibinfo {pages} {5188} (\bibinfo {year} {1976})}\BibitemShut {NoStop}%
\bibitem [{\citenamefont {Moon}\ and\ \citenamefont {Koshino}(2013)}]{Moon2013optical}%
  \BibitemOpen
  \bibfield  {author} {\bibinfo {author} {\bibfnamefont {P.}~\bibnamefont {Moon}}\ and\ \bibinfo {author} {\bibfnamefont {M.}~\bibnamefont {Koshino}},\ }\href {https://doi.org/10.1103/PhysRevB.87.205404} {\bibfield  {journal} {\bibinfo  {journal} {Phys. Rev. B}\ }\textbf {\bibinfo {volume} {87}},\ \bibinfo {pages} {205404} (\bibinfo {year} {2013})}\BibitemShut {NoStop}%
\bibitem [{\citenamefont {Gamayun}\ \emph {et~al.}(2018)\citenamefont {Gamayun}, \citenamefont {Ostroukh}, \citenamefont {Gnezdilov}, \citenamefont {Adagideli},\ and\ \citenamefont {Beenakker}}]{Gamayun2018valley}%
  \BibitemOpen
  \bibfield  {author} {\bibinfo {author} {\bibfnamefont {O.~V.}\ \bibnamefont {Gamayun}}, \bibinfo {author} {\bibfnamefont {V.~P.}\ \bibnamefont {Ostroukh}}, \bibinfo {author} {\bibfnamefont {N.~V.}\ \bibnamefont {Gnezdilov}}, \bibinfo {author} {\bibfnamefont {Ä.}~\bibnamefont {Adagideli}},\ and\ \bibinfo {author} {\bibfnamefont {C.~W.~J.}\ \bibnamefont {Beenakker}},\ }\href {https://doi.org/10.1088/1367-2630/aaa7e5} {\bibfield  {journal} {\bibinfo  {journal} {New Journal of Physics}\ }\textbf {\bibinfo {volume} {20}},\ \bibinfo {pages} {023016} (\bibinfo {year} {2018})}\BibitemShut {NoStop}%
\bibitem [{\citenamefont {Herrera}\ and\ \citenamefont {Naumis}(2020)}]{Herrera2020dynamic}%
  \BibitemOpen
  \bibfield  {author} {\bibinfo {author} {\bibfnamefont {S.~A.}\ \bibnamefont {Herrera}}\ and\ \bibinfo {author} {\bibfnamefont {G.~G.}\ \bibnamefont {Naumis}},\ }\href {https://doi.org/10.1103/PhysRevB.102.205429} {\bibfield  {journal} {\bibinfo  {journal} {Phys. Rev. B}\ }\textbf {\bibinfo {volume} {102}},\ \bibinfo {pages} {205429} (\bibinfo {year} {2020})}\BibitemShut {NoStop}%
\bibitem [{\citenamefont {Black-Schaffer}\ and\ \citenamefont {Honerkamp}(2014)}]{BlackSchaffer2014Chiral}%
  \BibitemOpen
  \bibfield  {author} {\bibinfo {author} {\bibfnamefont {A.~M.}\ \bibnamefont {Black-Schaffer}}\ and\ \bibinfo {author} {\bibfnamefont {C.}~\bibnamefont {Honerkamp}},\ }\href {https://doi.org/10.1088/0953-8984/26/42/423201} {\bibfield  {journal} {\bibinfo  {journal} {J. Phys.: Condens. Matter}\ }\textbf {\bibinfo {volume} {26}},\ \bibinfo {pages} {423201} (\bibinfo {year} {2014})}\BibitemShut {NoStop}%
\bibitem [{\citenamefont {Ojaj{\"a}rvi}\ \emph {et~al.}(2024)\citenamefont {Ojaj{\"a}rvi}, \citenamefont {Chubukov}, \citenamefont {Lee}, \citenamefont {Garst},\ and\ \citenamefont {Schmalian}}]{Ojajarvi2024pairing}%
  \BibitemOpen
  \bibfield  {author} {\bibinfo {author} {\bibfnamefont {R.}~\bibnamefont {Ojaj{\"a}rvi}}, \bibinfo {author} {\bibfnamefont {A.~V.}\ \bibnamefont {Chubukov}}, \bibinfo {author} {\bibfnamefont {Y.-C.}\ \bibnamefont {Lee}}, \bibinfo {author} {\bibfnamefont {M.}~\bibnamefont {Garst}},\ and\ \bibinfo {author} {\bibfnamefont {J.}~\bibnamefont {Schmalian}},\ }\href {https://doi.org/10.1038/s41535-024-00717-4} {\bibfield  {journal} {\bibinfo  {journal} {npj Quantum Materials}\ }\textbf {\bibinfo {volume} {9}},\ \bibinfo {pages} {105} (\bibinfo {year} {2024})}\BibitemShut {NoStop}%
\bibitem [{\citenamefont {Zaarour}\ \emph {et~al.}(2023)\citenamefont {Zaarour}, \citenamefont {Malesys}, \citenamefont {Teyssandier}, \citenamefont {Cranney}, \citenamefont {Denys}, \citenamefont {Bubendorff}, \citenamefont {Florentin}, \citenamefont {Josien}, \citenamefont {Vonau}, \citenamefont {Aubel}, \citenamefont {Ouerghi}, \citenamefont {Bena},\ and\ \citenamefont {Simon}}]{Zaarour2023Flat}%
  \BibitemOpen
  \bibfield  {author} {\bibinfo {author} {\bibfnamefont {A.}~\bibnamefont {Zaarour}}, \bibinfo {author} {\bibfnamefont {V.}~\bibnamefont {Malesys}}, \bibinfo {author} {\bibfnamefont {J.}~\bibnamefont {Teyssandier}}, \bibinfo {author} {\bibfnamefont {M.}~\bibnamefont {Cranney}}, \bibinfo {author} {\bibfnamefont {E.}~\bibnamefont {Denys}}, \bibinfo {author} {\bibfnamefont {J.~L.}\ \bibnamefont {Bubendorff}}, \bibinfo {author} {\bibfnamefont {A.}~\bibnamefont {Florentin}}, \bibinfo {author} {\bibfnamefont {L.}~\bibnamefont {Josien}}, \bibinfo {author} {\bibfnamefont {F.}~\bibnamefont {Vonau}}, \bibinfo {author} {\bibfnamefont {D.}~\bibnamefont {Aubel}}, \bibinfo {author} {\bibfnamefont {A.}~\bibnamefont {Ouerghi}}, \bibinfo {author} {\bibfnamefont {C.}~\bibnamefont {Bena}},\ and\ \bibinfo {author} {\bibfnamefont {L.}~\bibnamefont {Simon}},\ }\href {https://doi.org/10.1103/PhysRevResearch.5.013099} {\bibfield  {journal} {\bibinfo  {journal} {Phys. Rev. Res.}\ }\textbf {\bibinfo {volume} {5}},\ \bibinfo {pages}
  {013099} (\bibinfo {year} {2023})}\BibitemShut {NoStop}%
\end{thebibliography}%
\end{document}